\documentclass[fp,a4,twocolumn]{jpsj3}

\usepackage{txfonts}
\usepackage{bm,graphicx,subfigure,enumerate,latexsym,comment}
\usepackage{grffile}
\usepackage[dvips]{epsfig}
\usepackage{amsmath,amssymb}
\usepackage{color}
\usepackage{ulem}

\renewcommand{\tilde}{\widetilde}
\newcommand{\ltsim}{\protect\raisebox{-0.5ex}{$\:\stackrel{\raisebox{-0.6ex}{$\textstyle <$}}{\sim}\:$}} 

\newcommand{\gtsim}{\protect\raisebox{-0.5ex}{$\:\stackrel{\raisebox{-0.6ex}{$\textstyle >$}}{\sim}\:$}} 

\def\a{\alpha}
\def\b{\beta}
\def\c{\gamma}
\def\d{\delta}
\def\D{\Delta}

\def\br{\boldsymbol{r}}

\def\s{{\sigma}}

\def\bF{\boldsymbol{F}}

\def\br{\boldsymbol{r}}

\def\bsigma{\boldsymbol{\sigma}}

\def\bR{{\boldsymbol{R}}}

\def\bkappa{{\boldsymbol{\kappa}}}

\def\bR{\boldsymbol{R}}
\def\ro{{\text{o}}}

\def\rB{{\text{B}}}

\def\re{{\text{e}}}

\def\rM{{\rm{M}}}

\def\ro{{\rm{o}}} 
\def\rp{{\rm{p}}}

\def\rs{{\rm{s}}}

\def\rw{{\rm{w}}}

\def\rL{{\text{L}}}
\def\rM{{\text{M}}}

\def\rp{{\text{p}}}

\def\rs{{\text{s}}}

\def\Re{{\sf Re}}

\def\Dt{{\Delta t}}
\def\Dx{{\Delta x}}

\def\tkappa{{\tilde{\kappa}}}

\def\tsigma{{\tilde{\sigma}}}
\def\tt{{\tilde{t}}}
\def\tx{{\tilde{x}}}

\def\De{{\sf De}} 
\def\Dr{{\sf Dr}} 
\def\Re{{\sf Re}} 
\makeatletter
  \def\mathcomposite{%
     \@ifstar
        {\def\@mathcomposite@option{%
            \baselineskip\z@skip\lineskiplimit-\maxdimen}%
         \@mathcomposite}%
        {\let\@mathcomposite@option\offinterlineskip
         \@mathcomposite}}
  \def\@mathcomposite{%
     \@ifnextchar[\@@mathcomposite{\@@mathcomposite[0]}}
  \def\@@mathcomposite[#1]#2#3#4{%
     #2{\mathchoice
        {\@mathcomposite@{#1}{#3}{#4}\displaystyle{1}}%
        {\@mathcomposite@{#1}{#3}{#4}\textstyle{1}}%
        {\@mathcomposite@{#1}{#3}{#4}%
         \scriptstyle\defaultscriptratio}%
        {\@mathcomposite@{#1}{#3}{#4}%
         \scriptscriptstyle\defaultscriptscriptratio}}}
  \def\@mathcomposite@#1#2#3#4#5{%
     \vcenter{\m@th\@mathcomposite@option
        \dimen@\f@size\p@\dimen@#1\dimen@\dimen@#5\dimen@
        \divide\dimen@ 18
        \edef\@mathcomposite@skipamount{\the\dimen@}%
        \ialign{\hfil$#4##$\hfil\cr
           #2\crcr
           \noalign{\vskip\@mathcomposite@skipamount}%
           #3\crcr}}}
\makeatother

\makeatletter
\def\fgcaption{\def\@captype{figure}\caption}
\makeatother

\makeatletter
\def\fgcaption{\def\@captype{figure}\caption}
\makeatother

\makeatletter
\def\fgcaption{\def\@captype{figure}\caption}
\makeatother

\makeatletter
\def\fgcaption{\def\@captype{figure}\caption}
\makeatother

\title{Multiscale simulation of polymer melt spinning %
       by using the dumbbell model}

\author{
Takeshi Sato,
Kazuhiro Takase  
and
Takashi Taniguchi\thanks{E-mail: taniguch@cheme.kyoto-u.ac.jp}
}
\inst{
Department of Chemical Engineering, Kyoto University, Kyoto 615-8510, Japan 
}

\date{\today}

\abst{
We investigated the spinning process of a polymeric material
by using a multiscale simulation method which connects the
macroscopic and microscopic states through the stress and strain-rate
tensor fields, by using Lagrangian particles (filled with polymer chains) along the spinning line.
We introduce a large number of Lagrangian fluid particles
into the fluid, each containing $N_\rp$-Hookean-dumbbells to mimic the
polymer chains ($N_\rp$=10$^4$), which is equivalent to the upper
convected Maxwell fluid in the limit that $N_\rp$$\rightarrow$$\infty$.
Depending on the Reynolds number {\sf Re}, 
we studied the dynamical behaviors of fibers 
for the (a) {\sf Re}$\rightarrow$0
and (b) finite {\sf Re} cases,
for different draw ratios {\sf Dr}, ranging from 10 to 30
, and two typical Deborah numbers {\sf De}=10$^{-3} $
and {\sf De}=10$^{-2}$.
In the limit \Re$\rightarrow$0 (a),
as the Deborah number {\sf De} increases,
the elastic effect makes the system stable. 
At finite {\Re} 
(b), we found that 
inertial effects play an important role in determining 
the dynamical behavior of the spinning process, and for {\sf Dr}=10$^{-2}$
the system is quite stable, at least up to a draw ratio of {\sf Dr}=30. 
We also found that 
the fiber velocity and cross section area are determined 
solely by the draw ratio.
By comparing the velocity and cross section area profiles with the
end-point distribution for the dumbbell connective vectors, for
dumbbells located in Lagrangian particles along typical places along
the spinning line, we show that our multiscale simulation method
successfully bridges the microscopic state of the system with its
simultaneous macroscopic flow behavior.
It is also confirmed that the present schemes gives 
good agreements with the results obtained by the Maxwell constitutive equation. 
}

\kword{Multiscale simulation,
       Melt spinning process,
       Viscoelastic fluids,
       Maxwell model
       dumbbell model
       }

\begin{document}
\maketitle

\section{Introduction}

\medskip

\begin{figure}[h]
\begin{center}
\includegraphics[scale=1,width=7.0truecm]{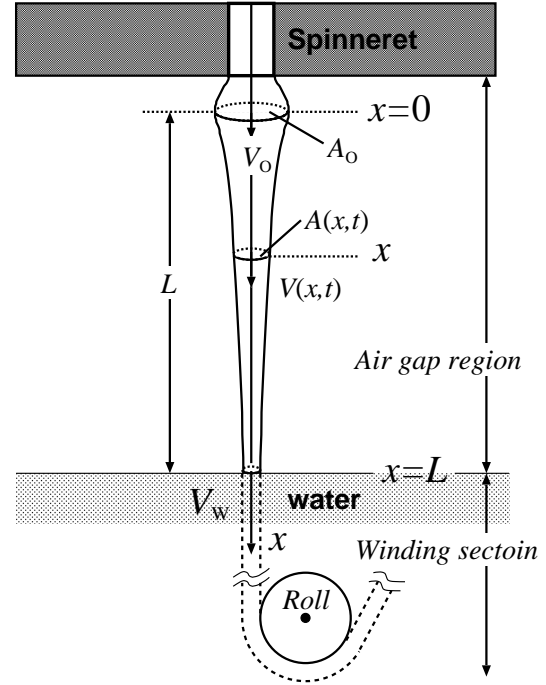}
\end{center}
\caption{Schematic view of the melt spinning process}
\label{fig:Fig01}
\end{figure}

Polymeric products have contributed 
to various industrial fields 
such as architecture, automotive, aerospace and medical fields.
To produce the desired products, 
it is important to develop appropriate polymer processing technologies,  
whereby one can control the specific properties of these products. 
One of the most common methods to manufacture a polymeric fiber 
is the melt-spinning process (see Fig.\ref{fig:Fig01})
\cite{
Ishihara_2011_1,
Ishihara_2011_2,
Ishihara_2011_3}. 
%
%
After the pioneering study 
of Kase and {Matsuo\cite{KaseMatsuo1965,KaseMatsuo1967}
for spinning processes of Newtonian fluids, 
a large number of theoretical and numerical studies
on the process have been performed by many researchers
\cite{%
MatovichPearson1969,
PearsonMatovich1969,
Gelder1971,
Denn1975,
Fisher1975,
Ishihara1976,
Ishihara_1989,
Ishihara_1992a,
Ishihara2006,
Takarada2001
}. 
%
%
At present, 
the simulation techniques of melt spinning processes
have been developed and applied to industrial problems,
but there still remains unsolved problems, if specifically say, 
predictions of the microscopic state of polymer chains
during spinning processes, 
such as the degree of orientation of polymer chains, 
position dependent entanglements density along a spinning line,
entanglement distribution on a single polymer chain 
and 
crystallization of semi-crystalline polymer on spinning process
\cite{Dhadwal2011}.
%
%
In most studies, a macroscopic approach using 
the Cauchy momentum equation with a constitutive equation 
to obtain the excess stress for the polymeric material
has been commonly used. 
However, it has been recognized that 
the stress obtained by such a constitutive equation
does not always predict the rheological properties 
of the target polymeric liquid correctly.
One may consider to simply use a molecular dynamics method 
to solve the macroscopic flow problem, however, the numerical cost of such an
approach is prohibitive.
To overcome this problem, 
we propose and develop 
a {\underline M}ulti{\underline S}cale {\underline S}imulation 
(MSS) method where the macroscopic model and microscopic molecular model are 
directly connected through the stress and strain rate tensor fields. 
This type of multiscale simulation method has been proposed 
by \"Ottinger with the concept of {``CONNFFESSIT''\cite{
Ottinger2005,
LasoOttinger1993
}}.
In the MSS method of the present work, 
we introduce Lagrangian particles 
as fluid elements, each of which contains 
many molecules to precisely describe the microscopic states.
So far,
for flow problems of polymeric fluids
a limited number of this type of multiscale simulations
have been performed
\cite{
Ottinger2005,
LasoOttinger1993,
Feigl1995,
WedgewoodGeurts1995,
HuaSchieber1996,
LasoPicassoOttinger1997,
vanHeelHulsenBrule1998,
HalinLielensKeuningsLegat1998,
BonvinPicasso1999,
RenE2005,
WeinanE2007,
YasudaYamamoto2008,
YasudaYamamoto2009,
YasudaYamamoto2010,
YasudaYamamoto2011,
MurashimaTaniguchi2010,
MurashimaTaniguchi2011,
MurashimaTaniguchi2011b,
Murashima2013
}.
Actually, the MSS method has been applied to solve 
flow problems of polymeric liquids 
in rather simple flow geometries 
such as flows in between two parallel plates{\cite
{
LasoOttinger1993,
WedgewoodGeurts1995,
HuaSchieber1996,
RenE2005,
WeinanE2007,
YasudaYamamoto2008,
YasudaYamamoto2009,
YasudaYamamoto2010,
YasudaYamamoto2011,
MurashimaTaniguchi2010,
Murashima2013
}}, 
flows around an infinitely long cylinder\cite{
vanHeelHulsenBrule1998,
BonvinPicasso1999,
MurashimaTaniguchi2011,
MurashimaTaniguchi2011b
}, 
flows in between eccentric rotating cylinders\cite{
LasoPicassoOttinger1997,
HalinLielensKeuningsLegat1998
}
and so forth. 
As far as we know, no one has ever attempted to apply this type of multiscale simulation method to an industrial polymer process before.
This means that an MSS method applicable 
to industrial polymer processing problems 
has not yet been established.
Hence, the aims of the present paper are
(i) to develop an MSS method applicable to a polymer melt spinning process
and (ii) to check the validity and efficiency of this method.
To make the assessment of the MSS method easier, 
in the present work 
we select a rather simple microscopic model for the polymer chains, 
{\it i.e.}, a Hookean dumbbell model, 
because it is well-known that 
the stress given by a set of Hookean dumbbells is equivalent 
to the one evaluated by the upper convected Maxwell constitutive equation,
provided the number of dumbbells goes infinity.
Although the dumbbell model is not a realistic polymer model,
we will be able to apply the present method 
to a more realistic situation just by replacing the model 
with a elaborated polymer model such as
{PASTA\cite{PASTA_Doi_Takimoto2001,Doi2003}} or
{NAPLES\cite{NAPLES_Masubuchi2001}}, among many others.

The content of this paper is as follows. 
In the next section, 
we explain our model for the melt spinning process 
at the macroscopic and microscopic levels separately. 
In Sec. \ref{sub:Equation of motion at macroscopic level}
we describe the governing equations at the macroscopic level, 
{\it e.g.}, the time evolution of the cross
section area and the fiber velocity, with the appropriate 
boundary conditions, and the assumptions we have used.
In Sec. \ref{sub:microscopic model of polymer chain and stress tensor}
we describe the set of equations used to investigate the dynamics of
the dumbbells in the microscopic model for the polymer chains.
In Sec. \ref{sub:Multiscale simulation method}
we explain how we have performed the multiscale simulations for a melt-spinning process 
as a function of the Reynolds number \Re.
In Sec. \ref{sec:Results of Multiscale simulations}
the results for the (a) {\sf Re}$\rightarrow$0 and
(b) finite {\sf Re} cases are presented separately, and explained in detail.
Finally, we give a summary in Sec. \ref{sec:Summary}.

\label{sec:Intro}

\medskip
\section{Melt spinning model}
\label{sec:Model of melt spinning}

   \subsection{Equations of motion at the macroscopic level}
    \label{sub:Equation of motion at macroscopic level}

   \medskip
   
Here we briefly explain the one dimensional model 
for a polymer melt spinning that we have used, even though such models
have been previously developed by other researchers
\cite{
Gelder1971,Shah1972,Ishihara1973,Fisher1975,Ishihara1976,Hyun1978b,Ishihara_1989,Yun2008}}.
A typical polymer melt spinning is drawn in Fig.\ref{fig:Fig01}.
A polymer melt is extruded from a die with a velocity $V_\ro$, 
and then the extrudate runs through the air.
%
%
Finally the filament is
taken up at a roll inside a water bath.
In deriving the set of equations for the one dimensional model
of a polymer melt spinning process, the following assumptions are made:
\begin{enumerate}
\renewcommand{\theenumi}{\roman{enumi}}
\renewcommand{\labelenumi}{(\roman{enumi})}
\item 
      The shape of the filament is axi-symmetric.
\item The polymer chains are relaxed at the place where 
      the diameter of the extruded filament from the orifice
      shows a maximum due to the die swelling effect.
      \label{item:die_swell}
\item The filament in the air gap region
      is isothermal\cite{Isothermal_Assumption}.
\item The gravitational force, 
      surface tension between the polymer melt and the air,
      and the friction of filament with the air are all neglected. 
\item The polymer filament is in the melt state just before reaching
      the surface of the cooling-water, and just after it has gone
      into the water bath, it is solidified instantaneously by cooling.
      \label{item:solidified}
\end{enumerate}
Under the assumptions mentioned above, 
the dynamics of the melt spinning process at the macroscopic level 
can be described by the cross section area $A(x,t)$, 
the velocity $V(x,t)$ 
and 
the tension $F(x,t)$ of the filament at a position $x$ and at time $t$.
The time evolution equation of the cross section area $A(x,t)$
is given by the following equation of continuity :
\begin{eqnarray}
{\partial A(x,t) \over \partial t} = - 
{\partial \over \partial x} \biggr ( A(x,t) V(x,t) \biggr ). 
\label{eqn:eq_of_continuity}
\end{eqnarray}	 
The tension of the fiber can be expressed as
\begin{eqnarray}
F(x,t) = A(x,t) \sigma(x,t),
\label{eqn:tension}
\end{eqnarray}
where $\sigma$ is the normal stress difference defined
by using the stress tensor of the polymer melt $\sigma_{\a\b}$
($\a,\b\in\{x,y,z\}$) as
\begin{eqnarray}
\sigma(x,t)=\sigma_{xx} - (\sigma_{yy}+\sigma_{zz})/2. 
\label{eqn:def_of_sigma}
\end{eqnarray}	 
The equation for the velocity 
is given by 
\begin{eqnarray}
\rho A { D V(x,t) \over D t } = {\partial \over \partial x} F(x,t), 
\label{eqn:force_balance}
\end{eqnarray}	 
where $D/Dt$=$\partial / \partial t$+$V \partial/\partial x$ 
is the Lagrangian derivative. 
%
%
%
%
%
%
%
Note that in an incompressible Newtonian fluid 
the uniaxial elongation stress $\sigma$ 
is expressed as $\sigma$=$3\eta_\rs {\partial V / \partial x}$, 
$\eta_\rs$ being the shear viscosity of the fluid. 
The boundary conditions for 
$A$ and $V$ at $x$=0 are given as 
\begin{eqnarray}
V(x=0,t)=V_\ro \quad \text{and}   
\quad  A(x=0,t)=A_\ro,  
\end{eqnarray}	 
where the position $x$=0 is defined 
such that the filament shows the maximum radius after 
being extruded from the die due to the Barus effect\cite{book:Bird1987}.
In addition, \
the normal stress difference 
$\sigma$ in eq.\eqref{eqn:def_of_sigma} is zero at $x$=0
because the polymer chains at $x$=0 are fully relaxed 
and isotropic, as given by assumption (\ref{item:die_swell}). 
In the spinning process, a tension is imparted to 
the polymer melt filament extruded from the die
by being taken up at the winding section at high speeds.
The filament is extended and becomes slender in the air gap region 
between the exit of the die and the water bath, and then it is cooled down 
by going into a cool medium, the water bath. 
As mentioned in assumption (\ref{item:solidified}),
because the filament is solidified just after entering the water bath 
(see Fig.\ref{fig:Fig01}),
the velocity of the filament at $x$=$L$ is the same as the winding velocity, 
$V_\rw$. 
Namely, the boundary condition for $V$ at $x$=$L$ is given as 
\begin{eqnarray}
V(x=L,t)=V_{\rw}.
\label{eqn:BoundaryCondition_to_V_at_x=1}
\end{eqnarray}	 
The velocity ratio between 
$V_\ro$ and the winding velocity $V_\rw$ 
is called the draw ratio, ${\sf Dr}$$\equiv$$V_\rw/V_\ro$. 
It is known that such elongational flow becomes unstable 
above a critical draw ratio ${\sf Dr}^{(\rm c)}$, 
where the draw resonance phenomenon appears and
the cross section area of the fiber varies periodically.
In addition, 
the air-gap distance,
temperature and
viscoelastic properties of the polymer melt
also affect the stability of the spinning process and 
the material properties of the resultant fibers\cite{
Ishihara_2011_1,
Ishihara_2011_2
}. 

All the variables defined at the macroscopic level
are scaled by using the spatial unit $\ell^{(\rM)}$=$L$,
time unit $t_\ro^{(\rM)}($=$L/V_\ro)$ and stress unit $\sigma_\ro^{(\rM)}$.
The symbols with the upperscript ${"(\rM)"}$ stands for
various units used at the \underline{M}acroscopic level,
and in the next subsection 
the symbols with the upperscript ${"(\rm m)"}$ are used for
units used at the \underline{m}icroscopic level.
In addition, the cross section $A$ is scaled by $A_\ro$.
%
%
%
%
The stress unit $\sigma_\ro^{(\rM)}$ will be explained in detail
in the next subsection,
after we introduce the corresponding constitutive equation.
%
%
All scaled variables are expressed with a tilde symbol on top, as
$\tilde V$=$V/V_\ro$, $\tilde A$=$A/A_\ro$,
$\tsigma$=$\sigma/\sigma_\ro^{(\rM)}$, 
$\tilde F$=$F / A_\ro\sigma_\ro^{(\rM)}$, $\tx$=$x/l_\ro^{(\rM)}$
and 
$\tt$=$t / t_\ro^{(\rM)}$.
After scaling, equations \eqref{eqn:eq_of_continuity}-\eqref{eqn:force_balance}
are found to be:
\begin{eqnarray}
{\partial \tilde A(\tx,\tt) \over \partial \tt}
&& \!\!\!\!\!  \!\!\!\!\!  \!\!\!
= - 
{\partial \over \partial \tilde x}
\biggr ( \tilde A(\tx,\tt) \tilde V(\tx,\tt) \biggr ), \qquad
\label{eqn:rescaled_eq_of_continuity}
\\
\tilde F(\tx,\tt)
&& \!\!\!\!\!  \!\!\!\!\!  \!\!\!
 = \tilde A(\tx,\tt) \tsigma(\tx,\tt),
\label{eqn:def_of_tension}
\\
\tsigma(\tx,\tt)
&& \!\!\!\!\!  \!\!\!\!\!  \!\!\!
 =\tsigma_{xx} - (\tsigma_{yy}+\tsigma_{zz})/2.
\label{eqn:def_of_rescaled_sigma}
\\
{\sf Re}~\tilde A
\biggr ( 
{\partial \tilde V \over  \partial \tilde t}
+ \tilde V { \partial \tilde V \over \partial \tilde x }
\biggr )
&& \!\!\!\!\!  \!\!\!\!\!  \!\!\!
= 
 {\partial \tilde F \over \partial \tx} 
\label{eqn:def_of_rescaled_momentum_equation}
\end{eqnarray}
where \Re=$\rho V_\ro^2/\sigma_\ro^{(\rM)}$ is the Reynolds number.
The boundary conditions are given as
\begin{eqnarray}
\tilde V(0,\tt) = 1,
\quad
\tilde V(1,\tt) = {\sf Dr}
\quad
\text{and}
\quad
\tilde A(0,\tt) = 1.
\label{eqn:rescaled_BoundaryCondition_to_V_at_x=1}
\end{eqnarray}
As seen from the set of equations 
\eqref{eqn:rescaled_eq_of_continuity}-\eqref{eqn:def_of_rescaled_momentum_equation}
and the boundary conditions \eqref{eqn:rescaled_BoundaryCondition_to_V_at_x=1},
the control parameters at the macroscopic level are
the draw ratio ${\sf Dr}$ and the Reynolds number {\sf Re}.

In industrial applications, 
the velocity of the spinning fiber 
is roughly categorized into four regions\cite{Ishihara_1992a,Ishihara_2011_1} :
(i) a low speed region  ($\ltsim$2000m/min),  
(ii) a partially oriented yarn region ($\sim$3500m/min), 
(iii) a high speed region             ($\sim$6000m/min)
and 
(iv) an ultra-high speed region ($>$6000 m/min).
The length of the air gap and 
the diameter of the nozzle are 0.05m$\ltsim L_\ro\ltsim$0.1m
(standard value $\sim$0.05m) 
and $A_\ro$$\simeq$0.2mm, respectively. 
If one considers a melt spinning process with a spinning velocity $V_\rw$
higher than that in the low speed region ($\ltsim$2000m/min), 
and uses $\eta_\rs$=10$^2$ ${\rm Pa} \cdot {\text s}$ 
as a typical value for the shear viscosity\cite{Ishihara_1992a} 
of the polymer melt used in the spinning, along with a
density $\rho_\ro$=10$^3{\rm kg/m}^3$, air gap $L_\ro$=0.05m  
and a typical stress of
$\sigma_\ro^{(\rM)}(=\eta_\rs V_\ro/L_\ro$), 
the Reynolds number {\sf Re} is found 
to be in the regime $ {\sf Re} \gtsim$16.7/{\sf Dr},
where the velocity of the fiber at the exit of the die 
is estimated as $V_\ro=V_\rw/{\sf Dr}$.
In the following sections, we separately investigate 
the two Reynolds number regimes
(a) \Re$\rightarrow$0 and
(b) finite \Re,     
for draw down ratios in the range 10$\le$\Dr$\le$30. 

%
%
%
%
%
%
%
%

   \subsection{Microscopic model for the polymer chains and stress tensor}
    \label{sub:microscopic model of polymer chain and stress tensor}

   \medskip
   
The rheological properties of the polymeric fluid 
play an essential role in determining 
the stability of the spinning process. 
Usually, a constitutive equation 
is used to theoretically predict the flow behavior. 
Until now, numerous constitutive equations have been proposed, 
and each of them presents its own advantages and disadvantages.
The main problems of using such a constitutive equation approach are:
(a)     how to choose a proper constitutive equation,
        which is able to predict the rheological properties of a
        target polymeric liquid,
        and
(b)     the difficulty in connecting the stress evaluated by 
        the constitutive equation
        to the microscopic state of of the polymer chains. 
In the present work, instead of using a macroscopic constitutive equation,
we employ a microscopic model to describe the rheology of the fluid, where
the stress tensor is evaluated in a statistical manner by considering the
microscopic states of the polymer chains.
So far, many microscopic polymer models have been proposed,
such as
the Rouse model\cite{Rouse1953},
the simple and finite extensible Hookean-dumbbell models\cite{book:Bird1987},
the Kremer-Grest-beads-spring model\cite{KremerGrest1990},
the Doi-Edwards reptation model\cite{DE1986}, 
the primitive chain network model\cite{NAPLES_Masubuchi2001},
and the slip-link models\cite{
ShanbhagLarson2001,
PASTA_Doi_Takimoto2001,
Doi2003,
Likhtman2005,
Schieber2003},
and so on, and they have all been used 
to investigate the bulk rheological properties 
of polymeric fluids. 
Here we employ the simple Hookean dumbbell model 
as a microscopic model in our multiscale simulation 
of the melt spinning process.
%
%
\begin{figure}[t]
\begin{center}
\includegraphics[scale=1,width=8truecm]{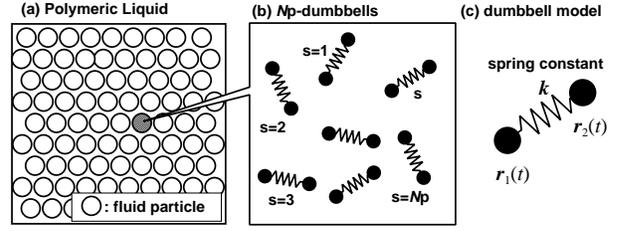}
\end{center}
\caption{
Schematic views of (a) fluid particles 
composing a polymeric liquid, 
(b) $N_\rp$-dumbbells in a fluid element 
and (c) the dumbbell model. 
}
\label{fig:Fig_02}
\end{figure}
%
%
The reasons why we have selected this particular model are the following: 
\begin{enumerate}
\renewcommand{\labelenumi}{(\roman{enumi})}
\item In the limit when the number of Hookean dumbbells goes to
  infinity, the corresponding constitutive equation is known. This
  allows us to assess the accuracy of our multiscale simulation. 
\item The computational cost is significantly lower than the other
      microscopic polymer models.
\end{enumerate}
%
%
%
%
The Hookean dumbbell is composed of two beads 
connected by a spring with a spring constant $k$. 
We consider that a fluid particle with a volume $v_\ro$
in a fluid contains $N_\rp$-dumbbells 
as shown in Fig.\ref{fig:Fig_02}(a) and (b).
It is assumed that no interaction does work among different dumbbells. 
The equation for the relative vector $\br$
between two beads is given as\cite{book:Bird1987}
\begin{equation}
{d\br \over dt} = \bkappa \cdot \br - { 2 \over \zeta } k \br 
  + { 1 \over \zeta } \bR
\label{eqn:r}
\end{equation}
where
$\zeta$ is the frictional coefficient,
$\bkappa$ is the velocity gradient tensor which is defined by 
$\kappa_{\a\b}$=${\partial V_\a / \partial x_\b }$ ($\a,\b\in\{x,y,z\}$) 
and $\bR$ is a random force
which satisfies the fluctuation-dissipation theorem. 
%
%
Using the obtained relative vectors $\br$ from eq.\eqref{eqn:r}
and
the spring forces $\bF$ for all the dumbbells,  
the stress tensor in a fluid particle can be evaluated by 
the following statistical average 
over constituent {dumbbells\cite{book:Bird1987},} 
which is known as Kramers' relation:
\begin{equation}
\sigma_{\a\b} =
  - {N_\rp \over v_\ro} \langle  r_{\a} F_{\b} \rangle
   ~\simeq~
  - { n \over N_\rp}
\sum_{s=1}^{N_\rp} r^{(s)}_{\a}F^{(s)}_{\b} ,  
\label{eqn:sigma_micro}
\end{equation}
where
$\langle \cdots \rangle$ denotes the statistical average of $(\cdots)$
and can be evaluated approximately by the average over $N_\rp$-dumbbells,
the variables with the upperscript $(s)$ mean the ones of the $s$-th dumbbell, 
$N_\rp$ is the number of dumbbells in the system of volume $v_\ro$ and 
$n$$\equiv$${N_\rp / v_\ro}$ is the number density of dumbbells. 
Using the expression of the spring force $\bF$=$-k \br$, 
the stress tensor is microscopically expressed as
\begin{equation}
\sigma_{\a\b}
= n \Bigr ( k \langle  r_\a r_\b \rangle -  k_\rB T \delta_{\a\b} \Bigr )
\label{eqn:microscopic_expression_of_stress}
\end{equation}
where
$k_\rB$ is the Boltzmann constant, $T$ the temperature,
and $\bsigma$ is redefined by subtracting a constant diagonal tensor from it,
such that it is equivalent to the zero tensor in the quiescent state,
by using the following relation 
$\langle r_\a r_\b \rangle_\text{eq}$=$k_\rB T \d_{\a\b}/k$.
%
%
%
%
Using eq.\eqref{eqn:microscopic_expression_of_stress},
the constitutive equation of the stress defined by 
\eqref{eqn:microscopic_expression_of_stress},
can be derived to be 
\begin{eqnarray}
{d\sigma_{\a\b} \over dt } =
 \kappa_{\a\c} \sigma_{\c\b}
+\sigma_{\a\c} \kappa_{\b\c}
+ nk_\rB T ( \kappa_{\a\b} + \kappa_{\b\a} )
- {4k \over \zeta } \sigma_{\a\b}, 
\label{eqn:constitutive_eq_for_dumbbell_model}
\end{eqnarray}
where the repeated indices are summed over $\{x,y,z\}$.
%
%
%
%
Equation \eqref{eqn:constitutive_eq_for_dumbbell_model}
is referred to as the upper convected Maxwell constitutive equation. 
It should be noted that
in the limit $N_\rp\rightarrow \infty$, 
the stress evaluated by eq.\eqref{eqn:microscopic_expression_of_stress}
converges to that evaluated 
by eq.\eqref{eqn:constitutive_eq_for_dumbbell_model}.
Because $N_\rp$ is always finite in a realistic numerical simulations,
the influence on rheological properties
of the statistical error always remains,
depending on how large $N_\rp$ is,
as shown in the last paragraph of this section. 

When constructing dimensionless expressions 
for eqs.\eqref{eqn:r}-\eqref
{eqn:constitutive_eq_for_dumbbell_model}
the Deborah number naturally appears.
The Deborah number $\sf De$ is the ratio
of an apparent convective time $t_\ro^{(\rM)}$(=$L/V_\ro)$ of a material point 
in the air gap region and 
the relaxation time of the dumbbell $\lambda$ 
$(\equiv$$\zeta/(4k))$, and is defined as \De=$\lambda/t_\ro^{(\rM)}$.
%
%
To obtain dimensionless expressions for
eqs.\eqref{eqn:r}-\eqref{eqn:constitutive_eq_for_dumbbell_model},
$t_\ro^{\rm (m)}$, $\ell_\ro^{\rm (m)}$ and $\sigma_\ro^{\rm (m)}$ 
are used as units of time, length and stress, respectively. 
We set the units at the microscopic level to be the same 
as those at the macroscopic level, except that 
the length of dumbbell is scaled by 
the equilibrium length of a dumbbell $\ell_{\rm eq}$=$\sqrt{3k_\rB
  T/k}$.
Namely, 
we chose as units of length and time $\ell_\ro^{\rm (m)}$=$\ell_\ro^{(\rM)}$=$L_\ro$ 
and ($t_\ro^{\rm (m)}$=$t_\ro^{(\rM)}$=$L/V_\ro$), respectively. 
We now introduce the unit of stress, which
we did not clearly explain when introducing the macroscopic
level model in eqs.\eqref{eqn:def_of_tension}-\eqref{eqn:def_of_rescaled_sigma},
since the constitutive equation had not yet appeared at that time. 
We define the units of stress as
$\sigma_\ro^{\rm (m)}$=$\sigma_\ro^{(\rM)}$=$n k_\rB T \lambda / t_\ro$.
It should be noted $\sigma_\ro^{\rm (m)}$ can also be expressed as
$\sigma_\ro^{\rm (m)}$=$\eta_\rs V_\ro / L_\ro$, with
$\eta_\rs$ being the shear viscosity of Maxwell fluid, given by $n k_\rB T \lambda$.
Hereafter $t_\ro$($\equiv$$t_\ro^\text{(m)}$=$t_\ro^\text{(M)})$ 
and $\sigma_\ro$($\equiv$$\sigma_\ro^\text{(m)}$=$\sigma_\ro^\text{(M)})$ 
will be used as units of time and stress.
Using the units defined above,
eqs.\eqref{eqn:r}-\eqref{eqn:constitutive_eq_for_dumbbell_model}
are rescaled and found to be
\begin{eqnarray}
   {d \tilde \br(\tilde t) \over d\tilde t}
= \tilde \bkappa \cdot \tilde \br(\tilde t) 
         -   { 1 \over 2{\sf De} }
	 \tilde \br(\tilde t) 
    + \sqrt{ { 1 \over 3{\sf De} }~}~
      \tilde \bR(\tilde t).
\qquad 
\label{eqn:eq_of_r_3}
\end{eqnarray}
where
$\tilde{\boldsymbol{\kappa}}=\bkappa t_\ro$, 
$\tilde \bR$ is defined by 
$\tilde \bR$=$\sqrt{t_\ro / 4 k_\rB T \zeta~}\bR$ 
and satisfies $\langle \tilde R_\a \rangle$=0
and $\langle \tilde R_\a(\tilde t)\tilde R_\b(\tilde t') \rangle$
=$\delta_{\a\b}\delta(\tilde t-\tilde t')$.
The dimensionless expression for the stress tensor is given as
\begin{eqnarray}
\tsigma_{\a\b} = { 3 \over {\sf De} }
\biggr (
\langle {\tilde r}_\a{\tilde r}_\b \rangle 
               - {1 \over 3} \delta_{\a\b}
\biggr ).
\label{eqn:dimensionless_sigma}
\end{eqnarray}
The dimensionless expression of 
eq.\eqref{eqn:constitutive_eq_for_dumbbell_model}
is found to be
\begin{eqnarray}
{D\tsigma_{\a\b} \over D\tt } =
 \tkappa_{\a\c} \tsigma_{\c\b}
+\tsigma_{\a\c} \tkappa_{\b\c}
+ {1 \over {\sf De}} ( \tkappa_{\a\b} + \tkappa_{\b\a} )
- {1 \over {\sf De}} \tsigma_{\a\b}. 
\label{eqn:dimensionless_constitutive_eq_for_dumbbell_model}
\end{eqnarray}
Unless otherwise stated, in what follows we omit the tilde symbols 
on dimensionless variables, except where it might cause confusion to
occur, in order to simplify the resulting expressions.
As seen from the dimensionless equations 
\eqref{eqn:eq_of_r_3}-\eqref{eqn:dimensionless_constitutive_eq_for_dumbbell_model}, 
{\sf De} is the only control parameter in the equations 
at the microscopic level.
Therefore, the control parameters in the present system
are the draw ratio ${\sf Dr}$, the Reynolds number {\sf Re}
and the Deborah number {\sf De}.

\begin{figure}[t]
\begin{center}
\includegraphics[scale=1,angle=-90,width=8.5truecm]{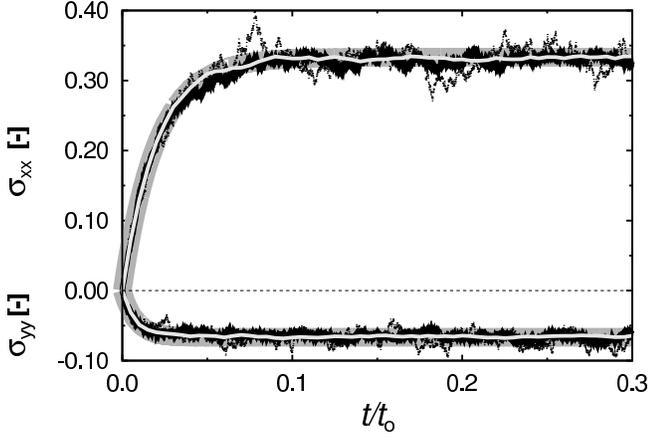}
\end{center}
\caption{
Time evolutions of $\s_{xx}$ and $\s_{yy}$ in a start-up flow
under a constant uniaxial elongational flow 
(${\kappa}_{xx}$=25) for Maxwell fluids ({\sf De}=10$^{-2}$)
consisting of non-interacting $N_\rp$-dumbbells 
($N_\rp$=10$^3$ (dotted-line), $10^4$ (black solid-line) and $10^5$ 
(white solid-line)).
The very thick grey line behind the other lines stands for
the theoretical lines obtained by 
the corresponding constitutive 
equation \eqref{eqn:constitutive_eq_for_dumbbell_model}.
}
\label{fig:Fig03}
\end{figure}

As mentioned in the previous paragraph, 
the statistical error of the stress tensor depends on 
the number of dumbbells $N_\rp$ used in the simulation.
Here we show how the stress evaluated 
by a microscopic system consisting of a finite number of $N_\rp$
deviates from the stress calculated 
by the mathematically equivalent constitutive equation 
\eqref{eqn:constitutive_eq_for_dumbbell_model},
which will give us useful information on how large $N_\rp$ should be
in order to obtain reasonable results when using 
multiscale simulations to describe melt spinning processes.
In Fig.\ref{fig:Fig03}, 
we show the time evolution of the stresses
in a uniaxial elongational start-up flow 
with a constant velocity gradient $\kappa_{xx}$=25
($\kappa_{yy}$=$\kappa_{zz}$=$-\kappa_{xx}$/2,
 $\kappa_{\a\b}$=0 for $\a$$\ne$$\b$)
for three different number of dumbbells, $N_\rp$=10$^3$, 10$^4$ and
10$^5$, with a Deborah number {\sf De}=10$^{-2}$.
The value $\kappa_{xx}$=25 is chosen as a typical velocity gradient
used in our multiscale simulations
and
\De=10$^{-2}$ is chosen so that the fluid
shows a viscoelastic behavior over a time period $t^\ast$,
during which a material point runs through the air gap region.
The time period $t^\ast$ is roughly estimated
to be 0.3$t_\ro$
by using $\Dr$$\simeq$$\kappa_{xx}$=25 and 
the expression $t^\ast$=$(\Dr-1)/(\Dr\ln\Dr$)
for the Newtonian fluid\cite{t_ast}.
%
%
%
%
As shown in Fig.\ref{fig:Fig03}, 
the deviation of the stress $\sigma_{xx}$ and $\sigma_{yy}$ 
from the theoretical lines obtained 
by using the upper convected Maxwell equation
\eqref{eqn:constitutive_eq_for_dumbbell_model}
decreases with increasing $N_\rp$.
We confirmed that
the standard deviation from $\sigma^{(\rM)}_{\a\a}$
decreases proportional to $1/\sqrt{N_\rp}$.
Judging from Fig.\ref{fig:Fig03} 
and from the view point of the trade-off between the accuracy of the stress 
and the computational cost, we decided to use $N_\rp$=$10^4$
as the number of dumbbells on a single Lagrangian particle 
({\it i.e.,} on each microscopic simulator) 
in the multiscale simulations presented in the next section.

   \subsection{Multiscale simulation method}
    \label{sub:Multiscale simulation method}

   \medskip
\begin{figure}[t]
\begin{center}
\includegraphics[scale=1,width=8.5truecm]{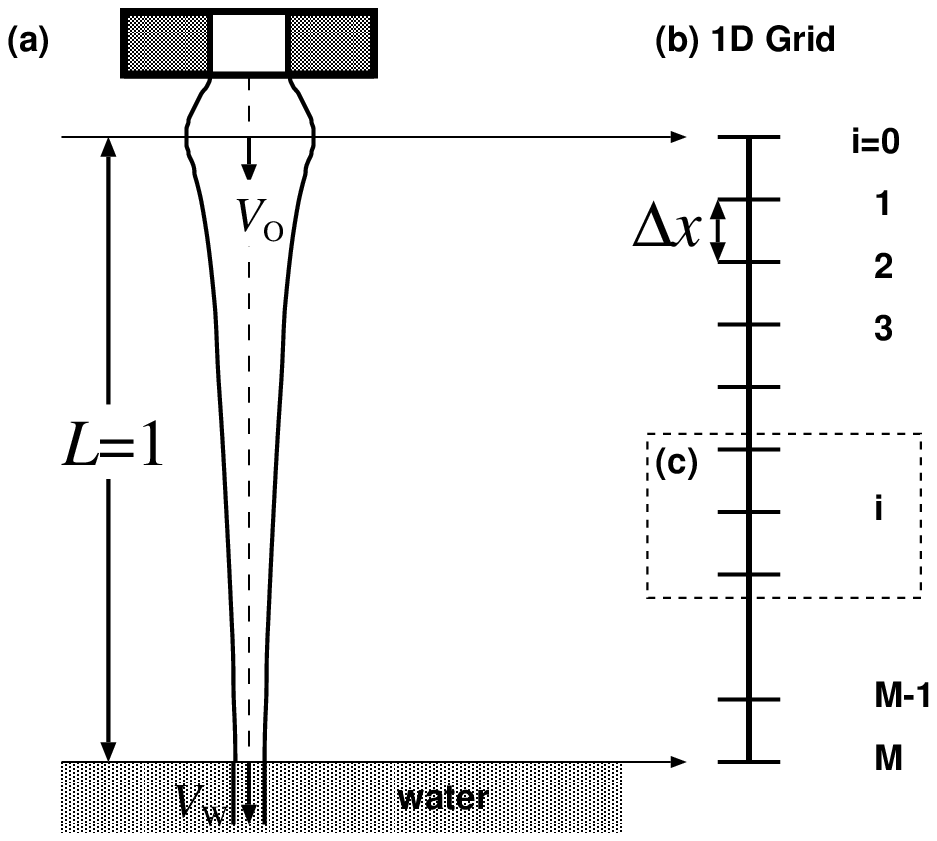}
\includegraphics[scale=1,width=8.0truecm]{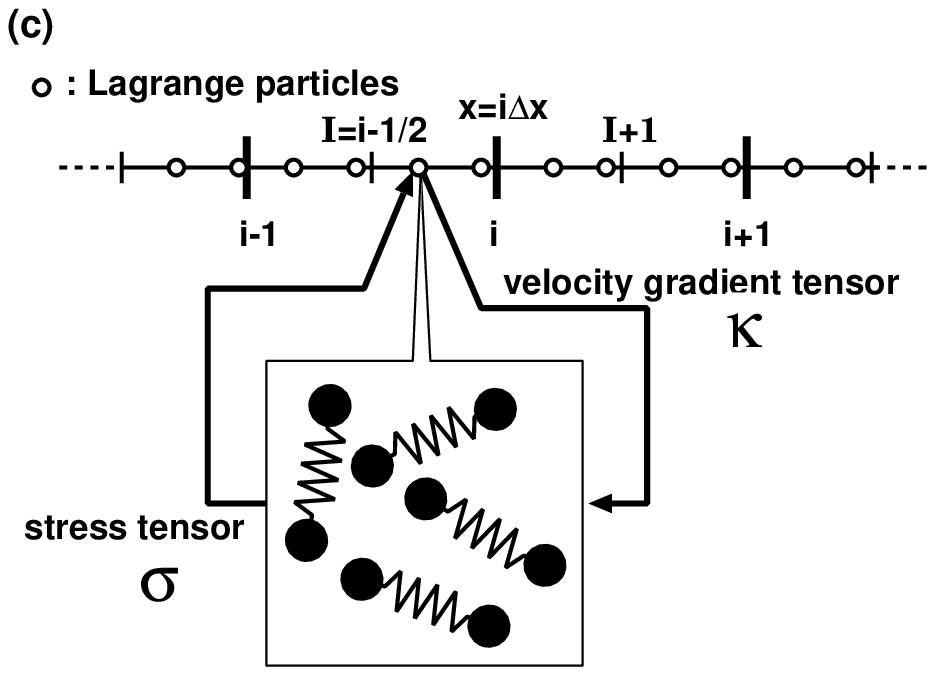}
\end{center}
\caption{Schematic views of (a) the filament in melt spinning process, 
(b) the grid used in solving the macroscopic equations,
and (c) the Lagrange fluid particles denoted by open circles
on the spinning line.}
\label{fig:Fig04}
\end{figure}
We performed numerical simulations of a spinning process 
in a multiscale way, where a set of macroscopic equations 
are solved by communicating with an ensemble 
of embedded microscopic simulation systems.
The fluid in the spinning process is
assumed to be a viscoelastic fluid described by a set of 
non-interacting Hookean dumbbells defined
in eq.\eqref{eqn:r},
which is equivalent to a Maxwell viscoelastic fluid, described by 
eq.\eqref{eqn:constitutive_eq_for_dumbbell_model}.
Because the stress of a viscoelastic fluid 
depends on the history of the strain and/or strain-rate 
that the microscopic molecules have experienced in the past, 
it is suitable to evaluate the stress tensor on a fluid particle 
based on the Lagrangian picture (see Fig.\ref{fig:Fig04}).
As shown in Fig.\ref{fig:Fig04}(b),
the spinning line in the air gap region 
is divided by ($M$+1)-grid points,
in other words, the line is divided into $M$ regions with 
a constant spatial interval $\D x$=$1/M$, 
to solve the equations at the macroscopic level
in the Eulerian fashion.
On a grid point $i$ (0$\le$$i$$\le$$M$),
the variables $A_i(t)$ and $V_i(t)$ are defined
($A_i(t)$$\equiv$$A(i \Delta x,t)$ and 
$V_i(t)\equiv V(i \Delta x,t)$), 
on the other hand, the stress $\sigma_I(t)$ is defined
on a staggered lattice point $I\equiv i-1/2$, 
($\sigma_I(t)\equiv \sigma(I\Delta x,t)$.
In addition, $N_\rL$-Lagrangian fluid particles 
are distributed on the spinning line, 
on each of which a microscopic simulation system is embedded. 
The stress tensor on a staggered lattice point $I$
is evaluated by an average over the stress tensors of all the Lagrangian particles
located within the spatial interval $[(i-1)\Dx,i\Dx]$. 

\medskip
\noindent
{(i) Cross section area update}

Suppose that the values of $A_i$ and $V_i$ at a time $t$ are known.
%
%
%
%
The cross section area $A_i(t)$
at a lattice point $i$ ($1$$\le$$i$$\le$$M$)
(see Fig.\ref{fig:Fig04}(b)) is updated 
by the following discretized form of equation 
of eq.\eqref{eqn:rescaled_eq_of_continuity}
\begin{eqnarray}
{A}_i(t+\D t) = A_i(t) + \delta A_i 
\label{eqn:discretized_eq_of_continuity}
\end{eqnarray}
where $\delta A_i$ is defined at time $t$ as
\begin{eqnarray}
\delta A_i = - (
 A_{i+1}V_{i+1}
-A_{i-1}V_{i-1}
 ) \D t / 2 \D x
\quad
\text{for}~~1 \le i < M
\end{eqnarray}
and 
$\delta A_M =
-( 3A_M V_M - 4A_{M-1}V_{M-1}+A_{M-2}V_{M-2} 
 ) \D t / 2 \D x $ 
with the boundary conditions $A_{0}(t)=1$ and $V_{0}(t)=1$
in eq.\eqref{eqn:rescaled_BoundaryCondition_to_V_at_x=1}. 

\medskip
\noindent
{(ii) Velocity field update}

The velocity $V_i(t+\Delta t)$
is evaluated from eq.\eqref{eqn:def_of_rescaled_momentum_equation}.
As we estimated at the end of Sec.~\ref{sub:microscopic model of
  polymer chain and stress tensor}, for moderate spinning 
the Reynolds number is in the range {\sf Re \gtsim 0.1}, 
but {\sf Re} may become very small
when we perform a very slow spinning or 
use a polymeric fluid with a very high viscosity. 
Because the appropriate numerical scheme for evaluating $V_i(t+\D t)$ 
depends on {\sf Re}, we separately explain the schemes used to update $V_i(t)$ 
for two cases,
(a) {\sf Re}$\rightarrow$0
and
(b) finite {\sf Re}.

\medskip
\noindent
(a) {\sf Re}$\rightarrow$0

For a very low Reynolds number, 
the force balance equation 
${\partial F / \partial x}$=0
will be satisfied instantaneously. 
Namely, at every single time step, 
the tension $F(I)$ must be balanced throughout the system. 
In the numerical simulation, the following condition 
should be satisfied:
\begin{equation}
{\delta F \over F_{\rm ave}} \equiv 
{~\bigr |~\text{Max}\{F(I)\} - \text{Min}\{F(I)\} ~\bigr |~
\over F_{\rm ave}
}
< \epsilon
\label{eqn:condition_for_F}
\end{equation}
where 
$\text{Max}\{F(I)\}$, 
$\text{Min}\{F(I)\}$
and 
$F_{\rm ave}$ 
denote the maximum, minimum and average of $F(I)$, respectively, and
$\epsilon$ is a suitably small value which sets the tolerance level.
To find the velocity $V_i(t+\Delta t)$ that satisfies 
condition \eqref{eqn:condition_for_F},
a relaxation method or a quasi-Newtonian method can be used.
In a simple relaxation method, 
the trial velocity $V_i^{(n)}$ at $t+\D t$
is repeatedly updated by using the following equation 
\begin{equation}
V^{(n+1)}_i = V^{(n)}_i + {\delta\over \D x} 
\bigr [ F^{(n)}(I+1)-F^{(n)}(I) \bigr ]
\label{eqn:relaxation_for_v}
\end{equation}
until the force balance condition given in eq.\eqref{eqn:condition_for_F},
with the updated stress values 
$\sigma_I^{(n+1)}(t+\Dt)$ obtained from the velocity gradient field
corresponding to $V^{(n+1)}_i$, is satisfied. In eq.\eqref{eqn:relaxation_for_v}, 
$\delta$ is a small virtual time increment,  
$n$ is the internal counter for the iteration step at $t+\D t$, 
$F^{(n)}(I)\equiv \sigma_I^{(n)}(t+\D t)A_I(t+\D t)$ 
and $A_I \equiv (A_{i-1}+A_i)/2$.
This means that at each step of the iterative procedure used to
find the $V_i(t+\D t)$ satisfying the force balance condition, 
the stress tensor $\sigma_I$ must be recalculated
according to the velocity gradient tensor corresponding to the trial $V_i^{(n)}$. 
As an alternative to this simple iterative procedure,
the Broyden method is one of the quasi-Newtonian methods 
used to find the roots of multi-dimensional non-linear
equations. This method can be very efficient, provided one gives an appropriate 
initial guess for $\{V_i(t+\D t)\}$, 
but if not, it can take many steps, or worse, it is possible that it
never converges.
%
%
%
%
%
%
%
%
Generally speaking, the relaxation method requires 
much more computational time than the Broyden method, 
but is relatively less-sensitive to the initial guess for $V_i$,
although a better initial guess will result in faster convergence.
However, for our present application, both methods required 
many internal iteration steps at each time step, 
and the velocity field sometimes would not converge 
within the specified threshold for the number of iteration steps, 
above which a simulation would not finish within a realistic time.
As will be shown in Sec.\ref{sec:Results of Multiscale simulations}{\sf (a)},
we chose the Broyden method to evaluate the velocity field
because it was relatively faster than the simple relaxation method.

\medskip
\noindent
(b) {\sf Finite Re} 

As estimated before,
for relatively high speed spinning ($\gtsim 2000$m/s),
the Reynolds number is in the range {\sf Re}$\gtsim$16.7/{\Dr}
({\rm e.g.}, 0.56 for \Dr=30).
In such situations, the contributions from the inertial term 
in eq.\eqref{eqn:def_of_rescaled_momentum_equation}
cannot be ignored. 
Therefore, we directly evaluate the velocity $V_i(t+\D t)$ 
by applying an explicit scheme for the momentum equation 
\eqref{eqn:def_of_rescaled_momentum_equation}
as 
\begin{eqnarray}
V_i(t+\Dt) = 
V_i(t) - \bigr (V_{i+1}(t))^2 - (V_{i-1}(t))^2 \bigr ) \D t /(4\D x) 
\nonumber\\
+ {\sf Re}^{-1} 
(\D t/\D x)
\big [ \sigma_{I+1}A_{I+1} - \sigma_{I}A_{I} \big ] / A_i. 
\label{eqn:scheme_to_solve_V_at_finite_Re}
\end{eqnarray}
%
%
%
%

\medskip
\noindent
{(iii) Lagrange particle update}

After obtaining the velocity $V_i$ on the Eulerian grids, 
as described above in the calculation step (ii),  
the velocities and velocity gradient tensors 
at all the positions of the Lagrange particles 
are evaluated by interpolating the $V_i$.
As an interpolating function for the velocity
and velocity gradient tensor $\kappa_{\a\b}$ we use
a quadratic and linear function, respectively.
Then, all the fluid particles are advected according 
to the local velocities evaluated at their positions. 
When the position $X(t)$ of a Lagrange particle closest to $x$=0
becomes larger than $\Delta X_{\rm init}$, 
a new Lagrangian particle is inserted into the system at position 
$X'$(=$X-\Delta X_{\rm init})$, 
in such a way that the initial distance between two adjacent Lagrange particles
is set to a constant $\Delta X_{\rm init}$.
Since the time interval $\D t$ used in our simulations is 
small,
the initial position $X'$ of the newly inserted Lagrangian particle
is almost zero.
In the simulations which will be shown in the next section,
the initial distance $\Delta X_{\rm init}$
is set to $\Delta X_{\rm init}$$\simeq$10$^{-4}$ in (a),
and 
$X_{\rm init}$=$10^{-4}$ for \Dr=10, 20 
and
$0.5\times 10^{-4}$ for \Dr=30 in (b).
The dumbbells on the newly inserted particle are 
set so as to satisfy the relation $<r_\a r_\b>=\delta_{\a\b}/3$,
which results in a zero stress tensor.

\medskip
\noindent
{(iv) Bridge between macro- and microscopic simulators}

A microscopic system embedded in a fluid particle
is used to obtain a local stress tensor as a response 
to the velocity gradient tensor 
at the position of a particle on the spinning line. 
The resulting stress tensors on the particles 
are transferred to the macroscopic system 
as stress tensors at the corresponding positions,  
then, the stress tensor $\sigma_I$ on the staggered lattice $I$ is evaluated 
as an average over the stresses on the particles   
located in the region-$I$ ({\it i.e.},
the spatial interval [$(i-1)\D \tx,i\D \tx$]).
Generally speaking, the step to evaluate 
the stress tensor on each Lagrangian particle
demands a tremendous computational expense,
because of the large number of molecular agents (dumbbells) present
in a single Lagrangian particle. 
Therefore, we have implemented a parallel computational code
by using MPI (Message Passing Interface),
specifically designed for the calculation of the stresses,  
which plays an essential role in reducing the computational time 
to a realistic value.
The number of cores used in the parallel computations is ranged from 64 to 96
in (a), is 24 in (b)\cite{CPU}.
In this way, we have performed the communication between macro- and microscopic 
systems.

\section{Results of the Multiscale simulations}
\label{sec:Results of Multiscale simulations}

   \medskip
   In the multiscale simulations, 
we used $M$=$100$ as the number of the grid points, with
$N_\rp$=$10^4$ in each microscopic system of Lagrangian material points,
as explained in the previous section. 
In the following, 
we present the results obtained in 
the typical cases of (a) {\sf Re} $\rightarrow 0$ and 
(b) finite {\sf Re}.
The initial configurations of the dumbbells
in each of the Lagrangian particles
in (a) and (b) are set to be those at the relaxed state,
{\it i.e.,} so as to give $\langle r_\a r_\b \rangle$=$\delta_{\a\b}/3.$
The initial values of the cross section area and velocity 
along the spinning line are those 
of the analytic solution for a Newtonian fluid 
at the specified {\sf Dr}.

\noindent
(a) {\sf Re}$\rightarrow$0

In the limit {\sf Re} $\rightarrow 0$, 
the tension will be balanced within an infinitely small time duration, 
therefore, the velocity field must be determined 
so as to satisfy $\partial F/\partial x$=0.
To obtain a velocity field which satisfies this constraint,
we used the Broyden method mentioned in the previous section.
All the simulations in (a),
the time increment $\D t$ is set to $10^{-4}$.
%
%
\begin{figure}[t]
\begin{center}
\includegraphics[scale=1,width=5.6truecm,angle=-90]{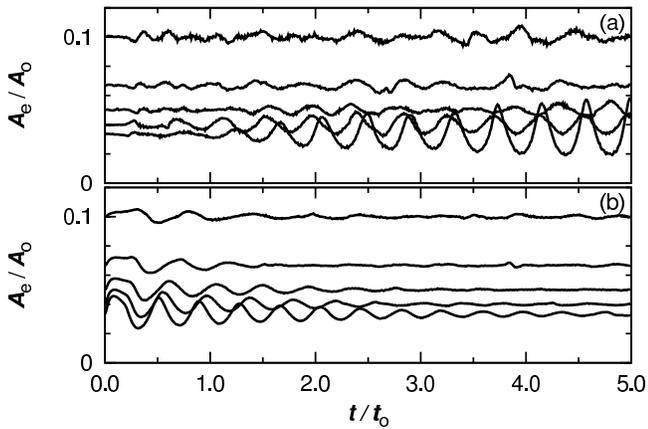}
\end{center}
\caption{
Time evolution of the cross section area $A_\re$ at an end point ($x$=1)
for {\sf Dr}=10, 15, 20, 25, 30 (from top to bottom) at 
(a) {\sf De}=$10^{-3}$
and 
(b) {\sf De}=$10^{-2}$.  
}
\label{fig:Fig05}
\end{figure}
Fig.\ref{fig:Fig05} 
shows the time evolutions of the cross section area $A_\re\equiv A(x$=$1,t)$
at the end of air gap region 
for {\sf Dr}=10, 15, 20, 25, 30 (from top to bottom)
at (a) {\sf De}=10$^{-3}$ and (b) {\sf De}=10$^{-2}$.
In the case of (a) {\sf De}=10$^{-3}$,  
the draw resonance phenomenon takes place at {\sf Dr}$\gtsim 20$. 
In the case of (b) {\sf De}=10$^{-2}$, 
although large oscillations of $A_\re(t)$ are observed at the
beginning, which increase with increasing values of {\sf Dr},
these oscillations are finally suppressed after $t\simeq 4$.
Namely, no periodic change in the cross section area 
could be observed in (b), 
although a fluctuation of $A_\re(t)$ still exists at long times.
Note that in the Newtonian fluid,
the critical draw ratio above which the draw resonance takes place 
is ${\sf (Dr)_c}$=$20.21$\cite{Gelder1971,Fisher1975}. 
It seems that 
$A_\re(t)$ for {\sf Dr}=20 and {\sf De}=10$^{-3}$
starts to show a periodic change. 
This result might be attributed to the effect of noise coming from 
the finite number of dumbbells we have used, but further
investigations should be performed in order to clarify such noise and
finite-number effects. 
It is considered that 
the suppression of the draw resonance in case (b) {\sf De}=10$^{-2}$
originates from the elastic effect of the dumbbells.
%
%

%
%
\begin{figure}[t]
\begin{center}
\includegraphics[scale=1,angle=-90,width=8.25truecm]{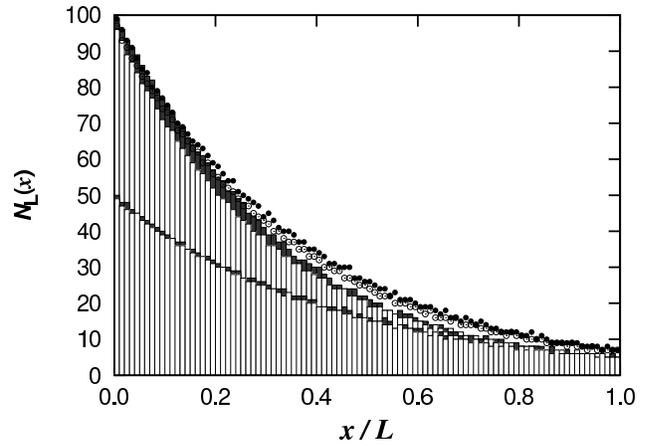}
\end{center}
\caption{
Spatial distributions of the number of Lagrangian particles $N_\rL(x)$
existing in the interval [$x,x+\D x$] on the spinning line 
for {\sf De}=$10^{-3}$ (grey) and {\sf De}=$10^{-2}$ (white) 
at $t$=5 in {\sf Dr}=10, 15 and 20.
The lower and higher histograms stand for {\sf Dr}=10 and 20, respectively,
but the results for {\sf Dr}=15 are drawn using filled circles
({\sf De}=$10^{-3}$) 
and open circles {\sf De}=$10^{-2}$,
to easily distinguish them from the other cases.
}
\label{fig:Fig06}
\end{figure}

Fig.\ref{fig:Fig06}
shows the spatial distribution of Lagrangian particles
existing in the region [$x,x+\D x$] with $\Delta x$=$1/M$,
at $t$=5, for {\sf Dr}=10, 15 and 20 
in {\sf De}=10$^{-3}$ (grey) and 10$^{-2}$ (white).
The lower and higher histograms, respectively,
denote the data for \Dr=10 and 20, 
and the results for {\sf Dr}=15 are drawn
using filled circles for \De=$10^{-3}$, 
and open circles for \De=$10^{-2}$.
When the position of the closest Lagrangian particle to $x$=0
becomes larger than a certain threshold value 
$\Delta X_{\rm init}$(=10$^{-4}$),
a new Lagrangian particle is inserted at a new position $(X$$\simeq$0), 
in such a way that the initial distance between 
the adjacent particles is constant $\Delta X_{\rm init}$.
Hence, the distribution of Lagrangian particles 
shows almost no change when the velocity field has reached a steady state. 
It should be noted that 
even when a draw resonance takes place,
the change in the velocity field is not so large 
that the particle distribution is significantly altered. 
The total numbers of Lagrangian particles $N_\rL^{\rm (Total)}$ 
for \Dr=10, 15 and 20 at $t$=5 are 
$N_\rL^{\rm (Total)}$$\simeq$ 
1900, 3300 and 3000, respectively,
and these numbers are almost constant for $t$$\gtsim$4.
%
%
To suppress the statistical error in evaluating the stress, 
it is important to have a large enough number of Lagrangian particles,
especially in the regions close to $x$=1,
because the density of particle becomes smallest at this position.
For all the cases considered here, 
we choose an appropriately small $\Delta X_{\rm init}$ 
so that the number of particles at $x\simeq 1$
is roughly constant,
which is enough to suppress the statistical error 
below a certain level.
Although a smaller $\Delta X_{\rm init}$ is better
for suppressing the error, 
a larger number of total Lagrangian particles
will make the computational cost increase significantly.
%
%

\begin{figure}[t]
\begin{center}
\includegraphics[scale=1,width=5.6truecm,angle=-90]{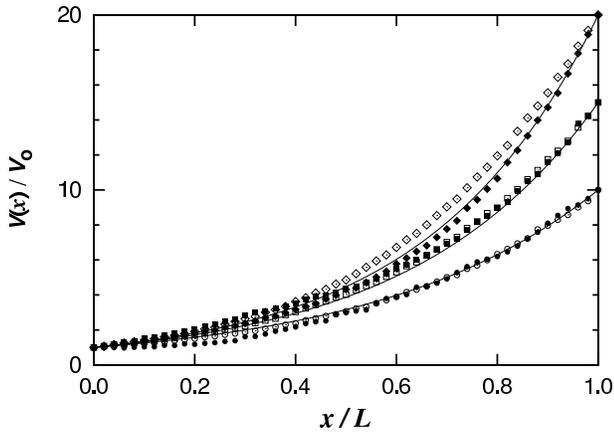}
\end{center}
\caption{
Velocity profiles at $t$=5
for {\sf Dr}=10 (circle), 15 (square) and 20 (diamond) 
on the spinning line obtained by the MSS method for 
{\sf De}=$10^{-3}$ (filled symbol) and $10^{-2}$ (open symbol),
The lines stand for the velocity profiles of Newtonian fluids 
for {\sf Dr}=10, 15 and 20, from bottom to top, respectively.
}
\label{fig:Fig07}
\end{figure}
Fig.\ref{fig:Fig07}
shows the velocity profiles on the spinning line at $t$=5 
for the six combinations of {\sf Dr}=10, 15, 20 and {\sf De}=10$^{-3}$, 10$^{-2}$.
For both {\sf De}=10$^{-2}$ and 10$^{-3}$,
the velocities for {\sf Dr}=10 and 15 
coincide with the corresponding ones of a Newtonian fluid (solid line)
given by eq.\eqref{eqn:AnalyticSolution_of_Newtonian_fluid} in Appendix, 
although a small deviation from the solid line can be seen
in the case of \De=10$^{-2}$ and {\sf Dr}=15.
We can infer from these results that  
the dumbbells in these cases are stretched,
which will be confirmed later in Fig.\ref{fig:Fig10}. 
At \Dr=20, 
the velocity profiles for {\sf De}=10$^{-3}$ 
mostly coincide with the Newtonian fluid results,
but the one for {\sf De}=10$^{-2}$ 
shows a clear deviation towards larger values.
This is because of the elastic effect 
coming from the slow relaxation of stretched dumbbells,
as seen from the analytic solution
(eq.\eqref{eqn:analytic_solution_in_elastic_limit} in Appendix)
in the elastic limit,
where $V(x)$ becomes a linear function of $x$.   
%
%
\begin{figure}[t]
\begin{center}
\includegraphics[scale=1,width=5.75truecm,angle=-90]{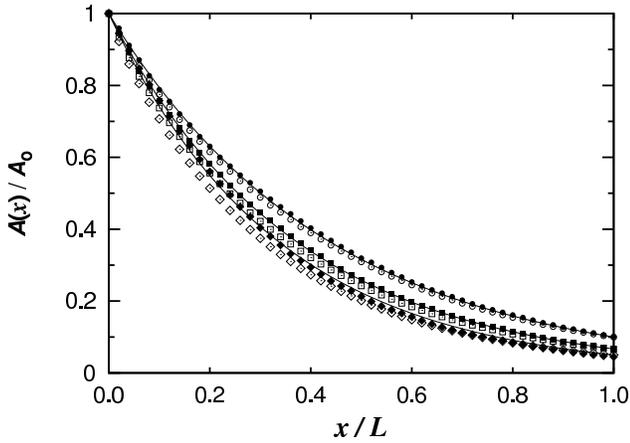}
\end{center}
\caption{
Cross section area $A(x)$ of the fiber obtained by the MSS method 
for {\sf Dr}=10 (circle), 15 (square) and 20 (diamond), and 
{\sf De}=$10^{-3}$ (filled symbols) and $10^{-2}$ (open symbols).
The solid lines are the cross section areas 
for a corresponding Newtonian fluid for {\sf Dr}=10, 15 and 20,
from top to bottom, respectively.
}
\label{fig:Fig08}
\end{figure}
In Fig.\ref{fig:Fig08}, 
we plot the cross section area along the spinning line 
for {\sf Dr}=10, 15 and 20 at {\sf De}=10$^{-3}$ and 10$^{-2}$.
The cross section areas at {\sf De}=10$^{-3}$
correspond almost exactly to the analytic solution for the Newtonian fluid. 
For the {\sf De}=10$^{-2}$ cases, on the other hand, 
the lines deviate towards smaller (thinner) values, with respect to
the corresponding Newtonian solution
(eq.\eqref{eqn:AnalyticSolution_of_Newtonian_fluid} in Appendix).
This tendency for $A$ is consistent with what would be expected given the 
velocity profiles shown in Fig.\ref{fig:Fig07}.
We can see that the $A_\re$ are mainly 
determined by {\sf Dr}, but they do not depend strongly on {\sf De}.
The state of the dumbbells at $x$$\simeq$1 for different \De, but
for the same \Dr, however, are significantly different to each other,
as can be clearly seen in Fig.\ref{fig:Fig10}.
%
%

\begin{figure}[t]
\begin{center}
\includegraphics[scale=1,width=5.75truecm,angle=-90]{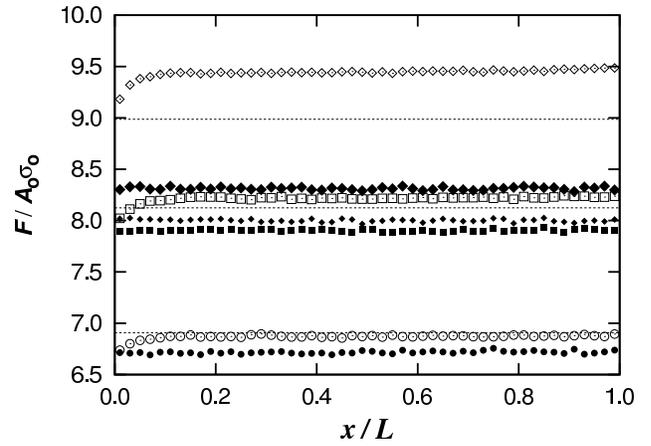}
\end{center}
\caption{
Tensions $F$ on the spinning line at $t$=5
for {\sf Dr}=10 (circle), 15 (square) and 20 (diamond), and 
{\sf De}=$10^{-3}$ (filled symbols) and $10^{-2}$ (open symbols).
As a reference, the tension at $t$=3
for {\sf Dr}=20 and {\sf De}=$10^{-3}$ 
is also shown by larger filled diamonds.
The dotted lines stand for tensions of a Newtonian fluid, 
$F$=$3\ln({\sf Dr})$ (See eq.\eqref{eqn:AnalyticSolution_of_Newtonian_fluid} in Appendix),
for {\sf Dr}=10, 15 and 20 from bottom to top, respectively. 
}
\label{fig:Fig09}
\end{figure}
%
%
\begin{table}[t]
\begin{center}
\caption{
Maximum ($F_{\rm max})$, minimum ($F_{\rm min})$, average ($F_{\rm
ave})$ and relative errors ($\delta F/F_{\rm ave})$ of the fiber tension
$F(x)$ at $t$=5 for (a) \Dr=10 (b) \Dr=15 (c) \Dr=20 (d) \Dr=25 (e) \Dr=30
at two typical Deborah numbers (\De=10$^{-3}$ and 10$^{-2}$),
where $\delta F$=$F_{\rm max}-F_{\rm min}$.  }
\begin{tabular}{llllll}
\hline
\hline
\parbox[l][14pt][c]{0cm}{}
\!\!\!\!
           ${\sf Dr}$
         & {\sf De}
         & $F_{\rm max}$ 
         & $F_{\rm min}$ 
         & $F_{\rm ave}$ 
         & $\delta F/F_{\rm ave}$
\\
\hline
\hline
\parbox[l][14pt][c]{0cm}{}
\!\!\!\!\!\!
          (a) 10
        & $10^{-3}$ 
        &  $2.25\times 10^{-3}$
        &  $2.23\times 10^{-3}$
        &  $2.24\times 10^{-3}$
        &  $1.05$\%
\\
\parbox[l][14pt][c]{0cm}{}
\!\!\!\!\!\!
           {\ }
        &  $10^{-2}$  
        &  $2.32\times 10^{-2}$
        &  $2.27\times 10^{-2}$
        &  $2.39\times 10^{-2}$
        &  $2.65$\%
\\
\hline
\hline
\parbox[l][14pt][c]{0cm}{}
\!\!\!\!\!\!
           (b) 15
        &  $10^{-3}$ 
        &  $2.26\times 10^{-3}$
        &  $2.26\times 10^{-3}$
        &  $2.63\times 10^{-3}$
        &  $0.58$\%
\\
\parbox[l][14pt][c]{0cm}{}
\!\!\!\!\!\!
           {\ }
        &  $10^{-2}$  
        &  $2.79\times 10^{-2}$
        &  $2.70\times 10^{-2}$
        &  $2.74\times 10^{-2}$
        &  $3.02$\%
\\
\hline
\hline
\parbox[l][14pt][c]{0cm}{}
\!\!\!\!\!\!
           (c) 20
        &  $10^{-3}$ 
        &  $2.27\times 10^{-3}$
        &  $2.27\times 10^{-3}$
        &  $2.67\times 10^{-3}$
        &  $0.73$\%
\\
\parbox[l][14pt][c]{0cm}{}
\!\!\!\!\!\!
           {\ }
        &  $10^{-2}$  
        &  $3.21\times 10^{-1}$
        &  $3.11\times 10^{-2}$
        &  $3.15\times 10^{-2}$
        &  $3.18$\%
\\
\hline
\hline
\parbox[l][14pt][c]{0cm}{}
\!\!\!\!\!\!
           (d) 25
        &  $10^{-3}$ 
        &  $3.34\times 10^{-3}$
        &  $3.31\times 10^{-3}$
        &  $3.32\times 10^{-3}$
        &  $0.89$\%
\\
\parbox[l][14pt][c]{0cm}{}
\!\!\!\!\!\!
           {\ }
        &  $10^{-2}$  
        &  $3.60\times 10^{-2}$
        &  $3.47\times 10^{-2}$
        &  $3.53\times 10^{-2}$
        &  $3.83$\%
\\
\hline
\hline
\parbox[l][14pt][c]{0cm}{}
\!\!\!\!\!\!
           (e) 30
        &  $10^{-3}$ 
        &  $4.52\times 10^{-3}$
        &  $4.45\times 10^{-3}$
        &  $4.50\times 10^{-3}$
        &  $1.70$\%
\\
\parbox[l][14pt][c]{0cm}{}
\!\!\!\!\!\!
           {\ }
        &  $10^{-2}$  
        &  $3.92\times 10^{-1}$
        &  $3.76\times 10^{-2}$
        &  $3.84\times 10^{-2}$
        &  $4.23$\%
\\
\hline
\hline
\label{tbl:Error_of_ForceBalance2}
\end{tabular}
\end{center}
\end{table}
As mentioned in the previous section, 
the tension along the fiber must be balanced. 
In our numerical simulations, this condition is 
relaxed by introducing a tolerance $\epsilon$, as 
explained in eq.\eqref{eqn:condition_for_F}, which
we have set to $\epsilon$=0.05.
In Fig.\ref{fig:Fig09}, 
the fiber tension along the spinning line is plotted, 
and we can clearly see that 
the tension is almost perfectly balanced.
In addition,
to understand to what extent the tension balance holds 
under the various conditions in our simulations, 
we present the maximum, minimum, and average 
tension at $t$=5 
along the spinning line, 
as well as the relative deviation of the tension from its average value
in Table \ref{tbl:Error_of_ForceBalance2}.
As seen from the table, 
the relative errors are maintained at less than 5\%,
as the tolerance $\epsilon$ in eq.\eqref{eqn:condition_for_F} 
is set to 0.05. 
As seen from Fig.\ref{fig:Fig09}, for a given {\sf Dr}, 
the tension at {\sf De}=10$^{-2}$ (open symbols)
is higher than the one at {\sf De}=10$^{-3}$ (corresponding filled symbols).
Although it is expected that the tension at {\sf De}=10$^{-3}$ 
(see filled symbols in Fig.\ref{fig:Fig09}) 
should be closer to the Newtonian values (shown as dashed lines),
the tension at $t$=5 for {\sf De}=10$^{-2}$ is actually closer. 
Particularly, the tension at {\sf De}=10$^{-3}$ for {\sf Dr}=20
is considerably smaller than
that of the corresponding Newtonian fluid
(see $F$ in eq.\eqref{eqn:AnalyticSolution_of_Newtonian_fluid}). 
We confirmed that the tension is changing largely 
with time depending on $A_\re$ in Fig.\ref{fig:Fig05}.
This is a reason why the tension at $t=5$ became smaller
than that of the Newtonian fluid. 
As a reference, 
the tension at $t$=3 for {\sf De}=10$^{-3}$ and {\sf Dr}=20
is also shown in Fig.\ref{fig:Fig09}, using larger filled diamonds.
The tension is apparently larger than that at $t$=5.
For the case of {\sf Dr}=20, $A_\re$ exhibits a periodic fluctuation,
with a minimum at $t$=5 (see Fig.~\ref{fig:Fig05}).
This implies that the tension is correlated to 
the change in $A_\re$, 
and therefore, that the smaller tension observed for \De=10$^{-3}$ and \Dr=20 
may be attributed to the larger fluctuations in the cross section area.
From the values given in Table \ref{tbl:Error_of_ForceBalance2},
we can see that 
the relative deviations in the tension, for the cases with 
{\sf De}=10$^{-3}$,
are significantly smaller than those with {\sf De}=10$^{-2}$.
As seen from Fig.\ref{fig:Fig09}, 
the maximum deviation in the tension for {\sf De}=10$^{-2}$
is located at the position closest to $x$=0, 
where the average configuration of the polymer chains 
is assumed to be isotropic, {\it i.e.,}
all the components of the stress tensor are zero. 
Thus, we infer that the deviation near $x$=0 comes 
from the assumed initial configurations of the dumbbells.
To obtain an initial configuration for the dumbbells 
without such an assumption, 
it is necessary to perform a multiscale simulation
even at the up-stream side,
including the die swell phenomenon,
as well as inside the die,
but this is outside the scope of the current work,
and will be investigated in a future work.
%
%
\begin{figure}[t]
\begin{center}
\includegraphics[scale=1,width=3.2truecm,angle=-90]{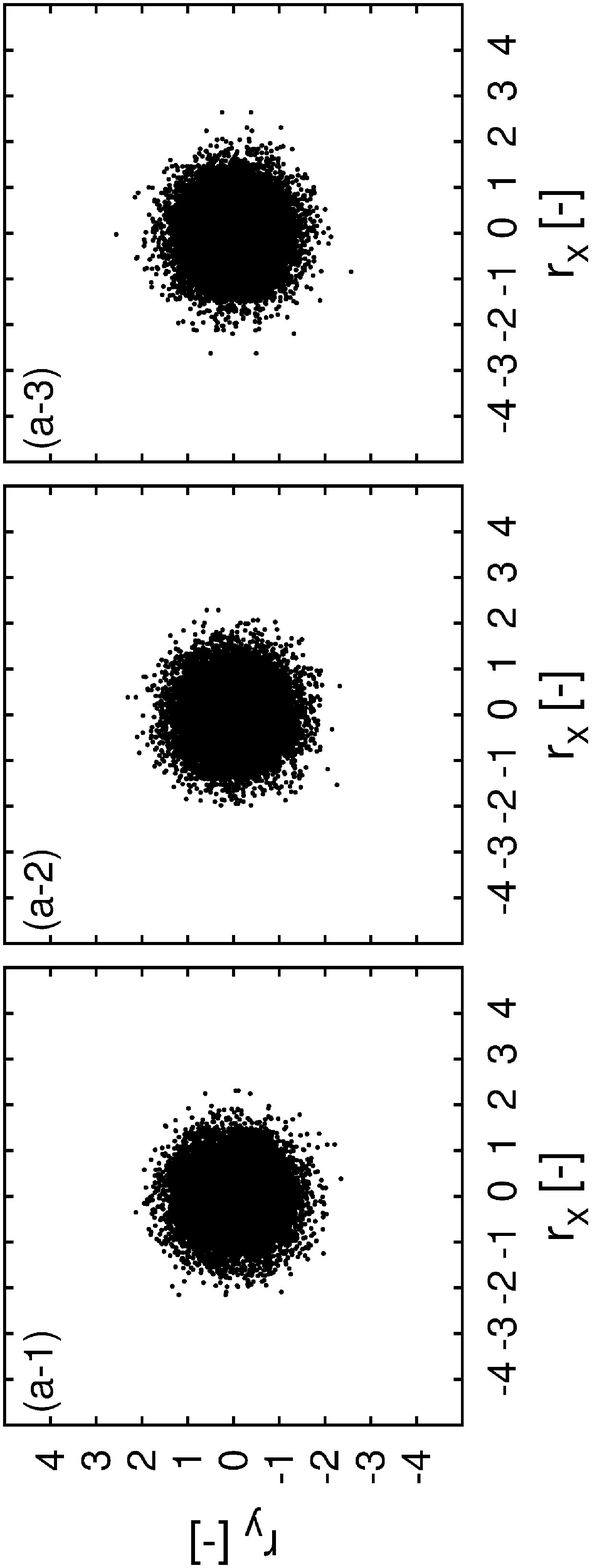}
\includegraphics[scale=1,width=3.2truecm,angle=-90]{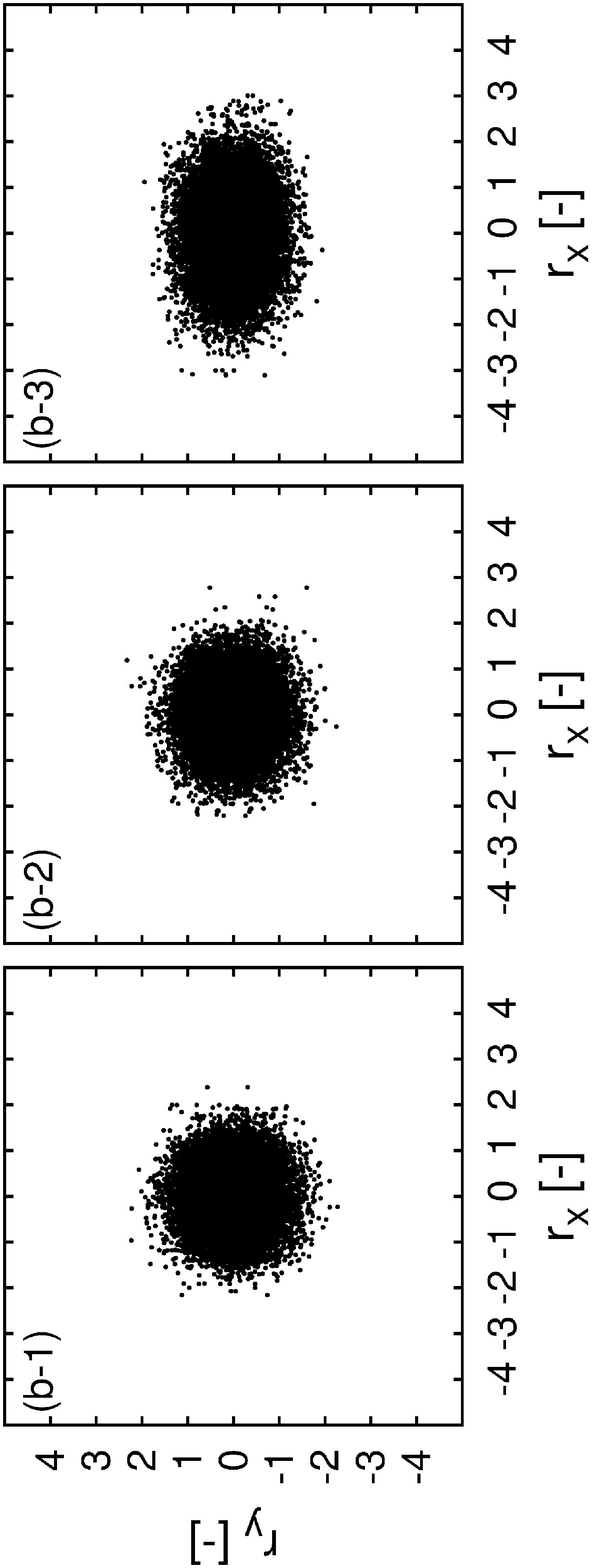}
\end{center}
\caption{
Distribution of the connecting vectors for \Dr=15
in a fluid particle at 
(1) $x$$\simeq$0.1, (2) $x$$\simeq$0.5 and (3) $x$$\simeq$1
at $t$=5 
for
(a) {\sf De}=10$^{-3}$ and 
(b) {\sf De}=10$^{-2}$. 
}
\label{fig:Fig10}
\end{figure}
%
%
\begin{figure}[t]
\begin{center}
\includegraphics[scale=1,width=3.2truecm,angle=-90]{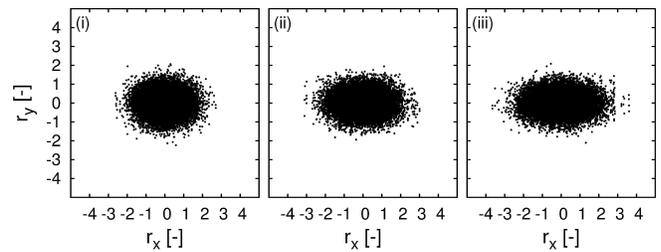}
\end{center}
\caption{
Distribution of connecting vectors
in a fluid particle located at $x$$\simeq$1 and at $t$=5 
for {\sf De}=10$^{-2}$
with
(i)  {\sf Dr}=10, 
(ii) {\sf Dr}=15 
and 
(iii) {\sf Dr}=20.
}
\label{fig:Fig11}
\end{figure}

Fig.\ref{fig:Fig10} shows 
the microscopic states of the dumbbells 
in Lagrangian particles located at three typical positions,  
(1) $x$$\simeq$0.1, (2) $x$$\simeq$0.5 and (3) $x$$\simeq$1 at $t$=5 
for (a) {\sf De}=10$^{-3}$ and (b) {\sf De}=10$^{-2}$ in {\sf Dr}=15.
In the figures, 
a dot represents the position of one of the end points 
of a dumbbell, with the other end point fixed at the origin,
where all of the end points within a single Lagrangian particle have been superimposed.
From the figures, one can see the degree of deformation of the dumbbells 
due to the uni-axial elongational flow along the spinning line.
In {\sf De}=10$^{-3}$ (Fig.\ref{fig:Fig10}(a)), 
the dots are almost isotropically distributed,
although the distribution in (a-3) is slightly 
elongated along the flow direction. 
In (b) {\sf De}=10$^{-2}$, on the other hand,
the distribution at (2) $x\simeq 0.5$ 
and (3) $x$$\simeq$1 is stretched along the $x$-direction,   
although the one at (1) $x$$\simeq$0.1 is almost isotropic. 
In particular,  in (b-3) one can see that 
the dumbbells are largely elongated in the $x$-direction
and shrunk in the $y$-direction.
Fig.\ref{fig:Fig11} shows 
the distribution of connecting vectors 
in a Lagrange particle located at $x$$\simeq$1 at $t$=5
for different {\Dr} 
with {\sf De}=10$^{-2}$.
%
%
From this Figure, 
we can see how the distribution is elongated 
as {\sf Dr} increases.
%
%
\begin{table}[t]
\begin{center}
\caption{
Conformation tensor $W_{\a\b}$ at $x\simeq$1 for
\De=$10^{-3}$ and \De=$10^{-2}$ 
for {\sf Dr}= (a) 10, (b) 15, (c) 20, (d) 25 and (e) 30.
}
\begin{tabular}{llllll}
\hline
\hline
\parbox[l][14pt][c]{0cm}{}
          ${\sf Dr}$
         & {\sf De}
         & $W_{xx}$ 
         & $W_{yy}$ 
         & $W_{xy}$
         & $-\sigma_{xx}/\sigma_{yy}$ 
\\
\hline
\hline
\parbox[l][14pt][c]{0cm}{}
            (a) 10
         &  $10^{-3}$ 
         &  $0.349$
         &  $0.326$
         &  \!\!\!\!$-2.89\times 10^{-5} $
         &  $2.15$
\\
\parbox[l][14pt][c]{0cm}{}
           {\ }
         & $10^{-2}$  
         &  $0.512$
         &  $0.281$
         &  $1.21\times 10^{-4} $
         &  $3.39$
\\
\hline
\hline
\parbox[l][14pt][c]{0cm}{}
            (b) 15
         &  $10^{-3}$ 
         &  $0.361$
         &  $0.321$
         &  $3.96\times 10^{-5} $
         &  $2.23$
\\
\parbox[l][14pt][c]{0cm}{}
           {\ }
         & $10^{-2}$  
         &  $0.675$
         &  $0.258$
         &  $9.55\times 10^{-5} $
         &  $4.51$
\\
\hline
\hline
\parbox[l][14pt][c]{0cm}{}
            (c) 20
         &  $10^{-3}$ 
         &  $0.374$
         &  $0.316$
         &  $2.43\times 10^{-5} $
         &  $2.35$
\\
\parbox[l][14pt][c]{0cm}{}
           {\ }
         & $10^{-2}$  
         &  $0.876$
         &  $0.240$
         &  \!\!\!$-4.06\times 10^{-5} $
         &  $5.82$
\\
\hline
\hline
\parbox[l][14pt][c]{0cm}{}
           (d) 25
        &  $10^{-3}$ 
        &  $0.390$
        &  $0.311$
        &  \!\!\!\!$-2.77\times 10^{-5} $
        &  $2.48$
\\
\parbox[l][14pt][c]{0cm}{}
           {\ }
        &  $10^{-2}$  
        &  $1.11$
        &  $0.227$
        &  \!\!\!\!$-1.07\times 10^{-5} $
        &  $7.32$
\\
\hline
\hline
\parbox[l][14pt][c]{0cm}{}
           (e) 30
        &  $10^{-3}$ 
        &  $0.408$
        &  $0.305$
        &  \!\!\!\!$-1.94\times 10^{-5} $
        &  $2.65$
\\
\parbox[l][14pt][c]{0cm}{}
           {\ }
        &  $10^{-2}$  
        &  $1.39$
        &  $0.216$
        &  \!\!\!\!$-7.26\times 10^{-5} $
        &  $9.07$
\\
\hline
\hline
\label{tbl:W_at_x=1}
\end{tabular}
\end{center}
\end{table}

Although we could qualitatively grasp the microscopic state
of the dumbbells at typical positions,
for various Deborah numbers and draw ratios, 
from Fig.\ref{fig:Fig10} and \ref{fig:Fig11},
in order to quantitatively understand the  microscopic states of the system, 
we focus on the statistically averaged quantities describing the state
of the dumbbells, 
such as their degree of stretching and orientation.
%
%
For this purpose, 
we have analyzed the conformation tensor $W_{\a\b}$ defined as 
\begin{equation}
W_{\a\b} =
{\langle r_\a r_\b \rangle}
= {1 \over N_\rp} \sum_{i=1}^{N_\rp} 
{r}_{i\a} 
{r}_{i\b}.
\end{equation}
%
%
It should be noted that this conformation tensor $W_{\a\b}$ 
is related to the scaled stress tensor $\sigma_{\a\b}$ 
of eq.\eqref{eqn:dimensionless_sigma}.
In Table \ref{tbl:W_at_x=1}, 
we give $W_{xx}$, $W_{yy}$ and $W_{xy}$ 
at a position in the vicinity of $x$=1
at $t$=5 
for ${\sf Dr}$=10, 15, 20, 25 and 30 and {\sf De}=10$^{-3}$
and 10$^{-2}$.
The values of $\sigma_{xx}/\sigma_{yy}$ at $x$$\simeq$$1$ 
are also shown in Table \ref{tbl:W_at_x=1}.
Given the axisymmetric geometry of the system, 
$W_{zz}$ is statistically equal to $W_{yy}$.
For the case of {\sf Dr}=10 and 15 at {\sf De}=10$^{-3}$, 
the $xx$- and $yy$-components are close to $1/3$,
and $xy$-component is nearly zero, 
which means the state of the dumbbells is almost relaxed and isotropic.
This result is consistent with Fig.\ref{fig:Fig10}(a-3).
In addition, 
$\sigma_{xx}/\sigma_{yy}$ in these cases is close to $-2$.
Because $\sigma_{xx}/\sigma_{yy}$=$-2$ holds 
in an axi-symmetric flow of an incompressible Newtonian fluid, 
the state of the fluid throughout the spinning line 
can be approximated by that of a Newtonian fluid, 
on the other hand, for the other cases,
the excess stress emerges.
%
%
\begin{figure}[t]
\begin{center}
\includegraphics[scale=1,width=5.6truecm,angle=-90]{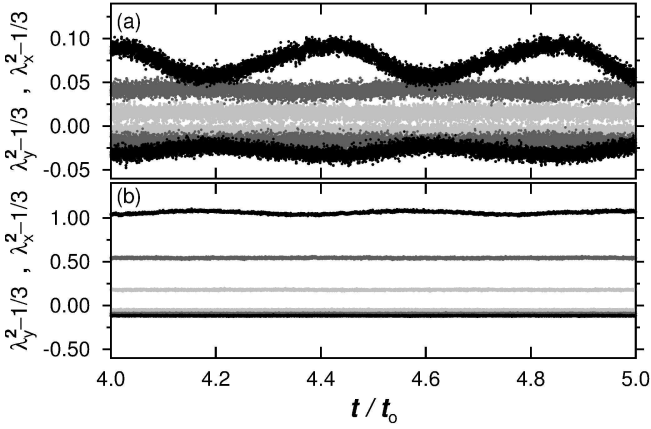}
\end{center}
\caption{
Time evolutions 
of  $\lambda_{x}^2-1/3$
and $\lambda_{y}^2-1/3$ at $x$=1 
for {\sf Dr}=10 (light grey), 15 (grey) and 20 (black)
at  (a) {\sf De }=10$^{-3}$
and (b) {\sf De }=10$^{-2}$.
$\lambda_{x}^2-1/3$ 
and $\lambda_{y}^2-1/3$ are positive and negative, respectively.
}
\label{fig:Fig12}
\end{figure}
\begin{figure}[t]
\begin{center}
\includegraphics[scale=1,width=5.6truecm,angle=-90]{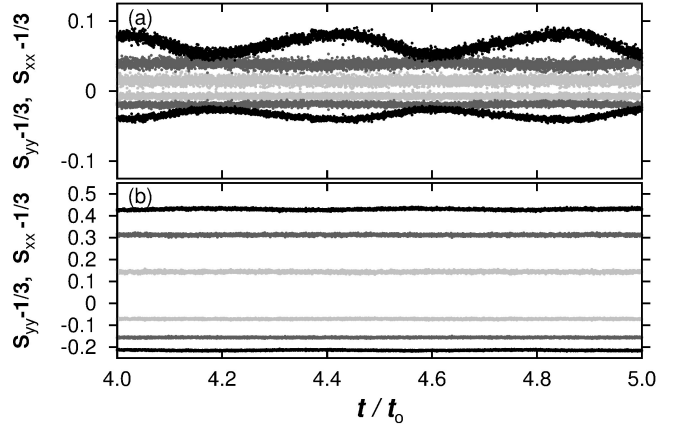}
\end{center}
\caption{
Time evolution of $S_{xx}$ and $S_{yy}$ 
for {\sf Dr}=10 (light grey), 15 (grey) and 20 (black)
in (a) {\sf De}=10$^{-3}$ and (b) {\sf De}=10$^{-2}$.
The  $xx$- and $yy$-components are positive 
and negative, respectively.
}
\label{fig:Fig13}
\end{figure}
%
%
The conformation tensor $W_{\a\b}$ 
can be decomposed into 
a stretch ratio $\lambda$ 
and an orientation tensor $S_{\a\b}$ as
\begin{equation}
W_{\a\b} = \lambda^2 S_{\a\b},
\label{eqn:expression_of_sigma}
\end{equation}
where $\lambda^2$ is defined as the mean squared length of the dumbbells :
\begin{eqnarray}
\lambda^2 = {\rm Tr}(\boldsymbol{W}) = \sum_{\a=x,y,z} \lambda_\a^2 
\label{eqn:lambda2}
\end{eqnarray}
with the stretch ratio in the $\a$-direction given by:
\begin{eqnarray}
\lambda_\a^2 = {1 \over N_\rp} \sum_{i=1}^{N_\rp} \br_{i\a}^2.
\end{eqnarray}
%
%
%
%
In Fig.\ref{fig:Fig12} and Fig.\ref{fig:Fig13},
we plot the time evolutions of the stretch ratio
along the $x$- and $y$-directions,
and the  $xx$- and $yy$-components orientation tensor, respectively,  
for {\sf Dr}=10, 15 and 20 at (a) {\sf De}=10$^{-3}$ and (b) {\sf De}=10$^{-2}$.
When the macroscopic cross section area becomes smaller, 
the dumbbells are stretched along the flow direction, but
shrunk in the perpendicular plane.
As such, the stretch ratio $\lambda_x^2-1/3$ is strictly positive,
and $\lambda_y^2-1/3$ negative.
In addition, the $xx$-component of the orientation tensor
is positive and the $yy$-component is negative for any parameter set,
as expected from Fig.\ref{fig:Fig03}.
As shown in Fig.\ref{fig:Fig12}, 
as the draw ratio becomes larger, 
the stretch ratio of the dumbbells along the flow direction $\lambda_{x}$
becomes larger.
We can see from Fig.\ref{fig:Fig12}-\ref{fig:Fig13},
that the stretch ratios, as well as the $xx$- and $yy$-components of
the orientation tensor, for {\sf Dr}=20 and {\sf De}=10$^{-3}$,
are temporally oscillating in phase with each other, 
and they are also synchronized with the change 
in $A_\re$ (for the corresponding parameter set shown in Fig.\ref{fig:Fig05}).
In the other cases, the stretch ratios and the orientation tensor show
almost constant values, although they are rapidly fluctuating. 
From all these results, 
we can conclude that 
the larger relaxation time results in a more stable melt spinning process.
\begin{figure}[t]
\begin{center}
\includegraphics[scale=1,width=5.6truecm,angle=-90]{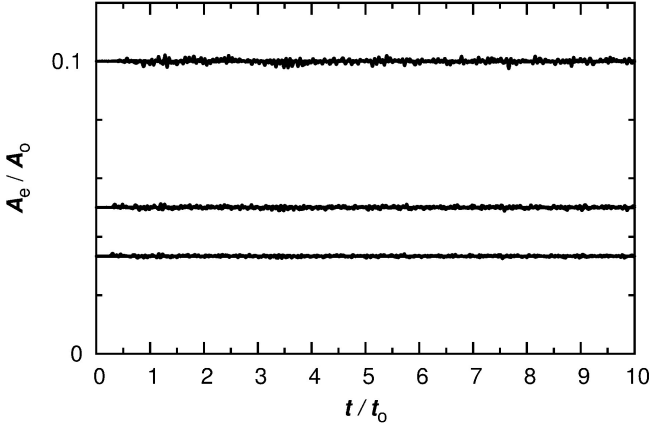}
\end{center}
\caption{
Time evolution of the cross section area $A_\re$ at the end point 
($x$$\simeq$1), for the eighteen combinations of  
{\sf Dr}=10, 20, 30,  {\sf Re}= 10$^{-2}$, 10$^{-1}$, 1 
and {\sf De}=10$^{-3}$, 10$^{-2}$. 
The results all converge into three lines, depending only on the draw ratio, 
\Dr=10 (top), 20 (middle), 30 (bottom).
}
\label{fig:Fig14}
\end{figure}

%
%
\medskip
\noindent
(b) {\sf Finite Re}

As explained at the end of 
Sec.\ref{sub:microscopic model of polymer chain and stress tensor},
the typical Reynolds number in industrial spinning processes 
easily reaches values larger than 10$^{-1}$. 
Because the inertial effect cannot be ignored 
when solving the momentum equation \eqref{eqn:def_of_rescaled_momentum_equation}
in this $\Re$-regime,
we use the scheme described in eq.\eqref{eqn:scheme_to_solve_V_at_finite_Re}
with $N_\rp$=10$^4$ dumbbells in each Lagrangian particle.
In (b) {\sf Finite Re}, the time increment $\D t$ is set to
$\Delta t$=$1.0\times 10^{-5}$ for \Re=1.0 and 0.1,
and
$\Delta t$=$5.0\times 10^{-6}$ or $\Delta t$=$1.0\times 10^{-6}$ for \Re=0.01.
Fig.\ref{fig:Fig14} shows 
the time evolution of the cross section area $A_\re(t)$ at $x$$\simeq$1.
Despite performing simulations for the eighteen different combinations
of the following parameters {\sf Dr}=10, 20, 30,  
{\sf Re}=10$^{-2}$, 10$^{-1}$, 1 and {\sf De}=$10^{-3}$, $10^{-2}$,
the resulting values for the relative cross section areas all converge
to three lines, depending only on the {\sf Dr}=10, 20, 30; with smaller
{\sf Dr} corresponding to larger $A_e$, as shown in in Fig.\ref{fig:Fig14}.
While the cases for \Dr$\gtsim$20 and {\sf De}=10$^{-3}$ at {\sf Re }$\ll$1 
exhibit the draw resonance phenomenon, 
the flow and cross section area in the present case 
are quite stable, even at {\sf Dr}=30 
for {\sf De}=10$^{-3}$.
We have found that the inertial effects make the spinning process stable,
in addition to the elastic effect.
%
%
\begin{figure}[t]
\begin{center}
\includegraphics[scale=1,width=6.0truecm,angle=-90]{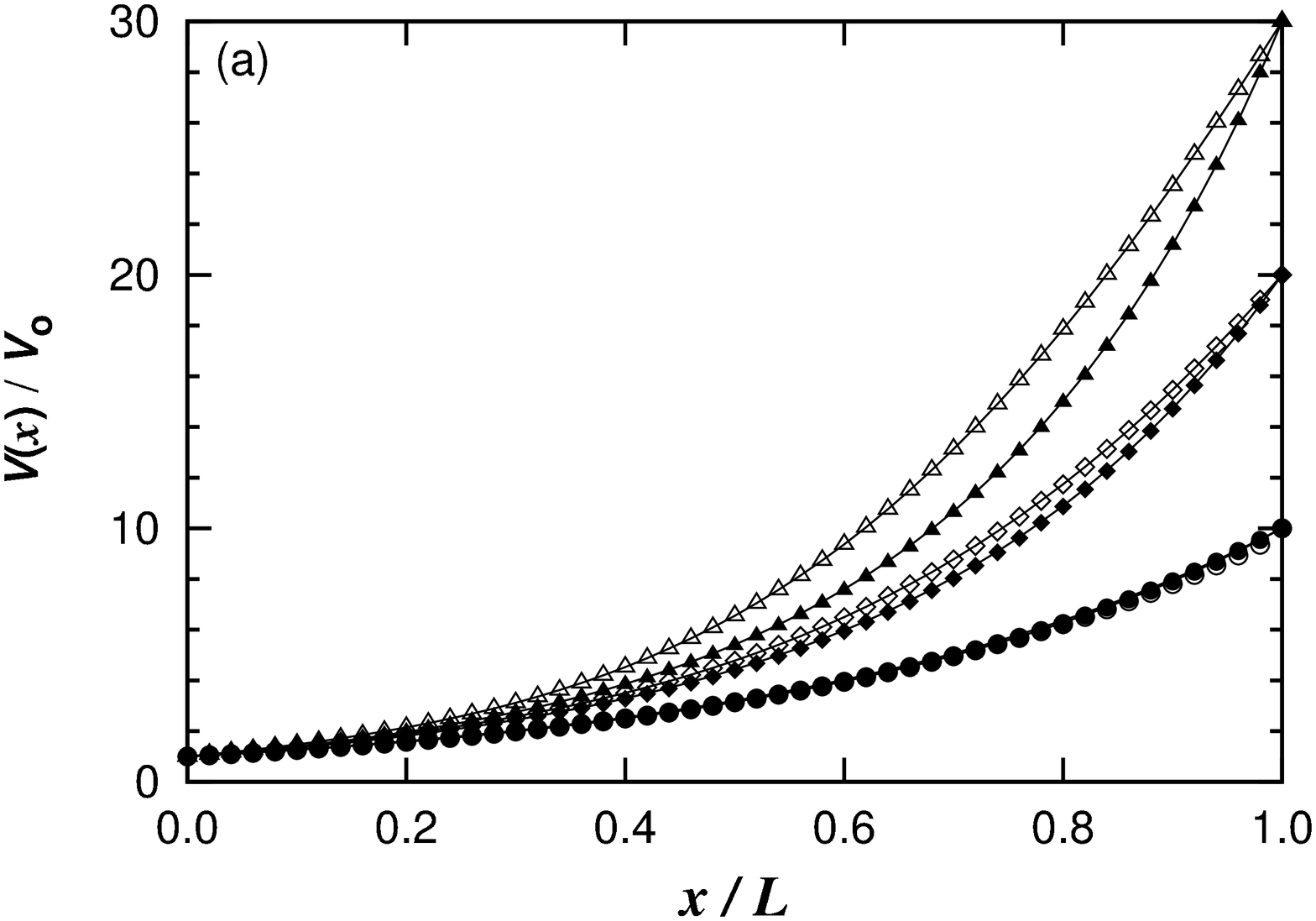}
\includegraphics[scale=1,width=6.0truecm,angle=-90]{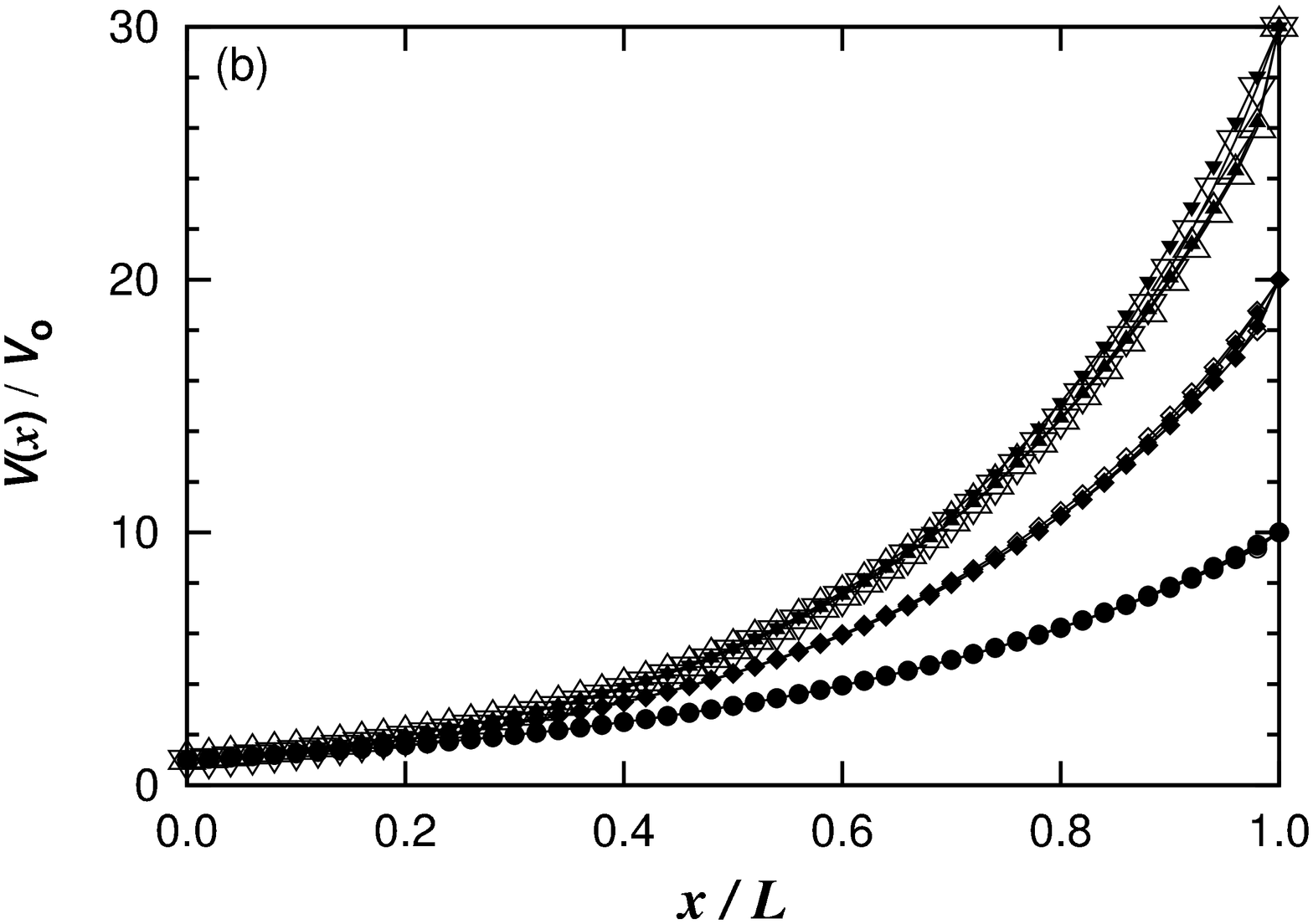}
\end{center}
\caption{
Velocity profiles at $t$=10 
for (a) {\sf Re}=10$^{-2}$ and (b) 10$^{-1}$ and 1.
The six combinations of the
three draw ratios {\sf Dr}=10 (circle), 20 (diamond), 30 (triangle) and
two Deborah numbers {\sf De}=10$^{-3}$ (filled symbol)
and 10$^{-2}$ (open symbol) are plotted for each \Re.
In (b), triangle symbols for \Re=10$^{-1}$ and \Re=1
are distinguished by using lower and upper triangles, respectively.
Namely, the velocity close to $x$$\simeq$1 for \Re=10$^{-1}$
is slightly higher than that for \Re=1.
The solid lines are the velocity profiles obtained 
by using the Maxwell constitutive equation 
with the parameters corresponding to the nearest symbols.
%
%
}
\label{fig:Fig15}
\end{figure}
%
%

In Fig.\ref{fig:Fig15},  
we separately show the velocity profiles
for (a) \Re=10$^{-2}$ and (b) 10$^{-1}$ and 1.  
For each Reynolds number, we plot 
the data obtained for the six following parameter combinations: 
{\sf Dr}=10 (circle), 20 (diamond), 30 (triangle), 
and {\sf De}=10$^{-3}$ (open symbols) and 10$^{-2}$ (filled symbols).
In addition, the results obtained by using
the Maxwell constitutive equation are plotted by solid lines. 
As seen from the figures, the velocity profiles for {\sf Dr}=10 
are irrespective of {\sf De} and {\sf Re}.
At {\sf Dr}=20 and 30, the velocity profiles
are almost the same, except 
for the case with {\sf Re}=10$^{-2}$ and {\sf De}=10$^{-3}$
(open triangles in (a))
whose velocity profile deviates towards higher values.
From these results,
the velocity profiles for {\sf Re}$\gtsim$10$^{-1}$ seem to depend
only on the draw ratio.
%
%
\begin{figure}[t]
\begin{center}
\includegraphics[scale=1,width=6.0truecm,angle=-90]{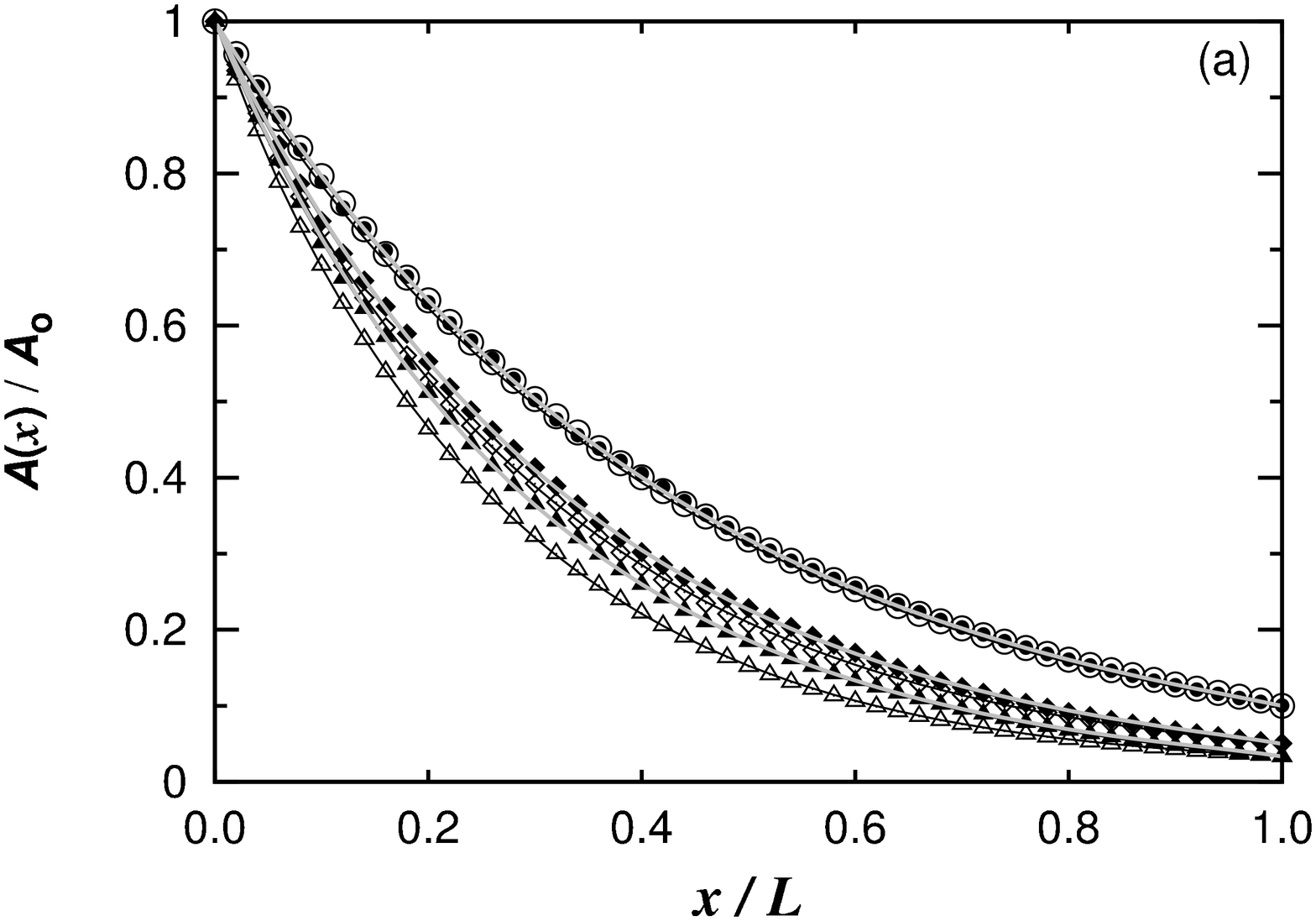}
\includegraphics[scale=1,width=6.0truecm,angle=-90]{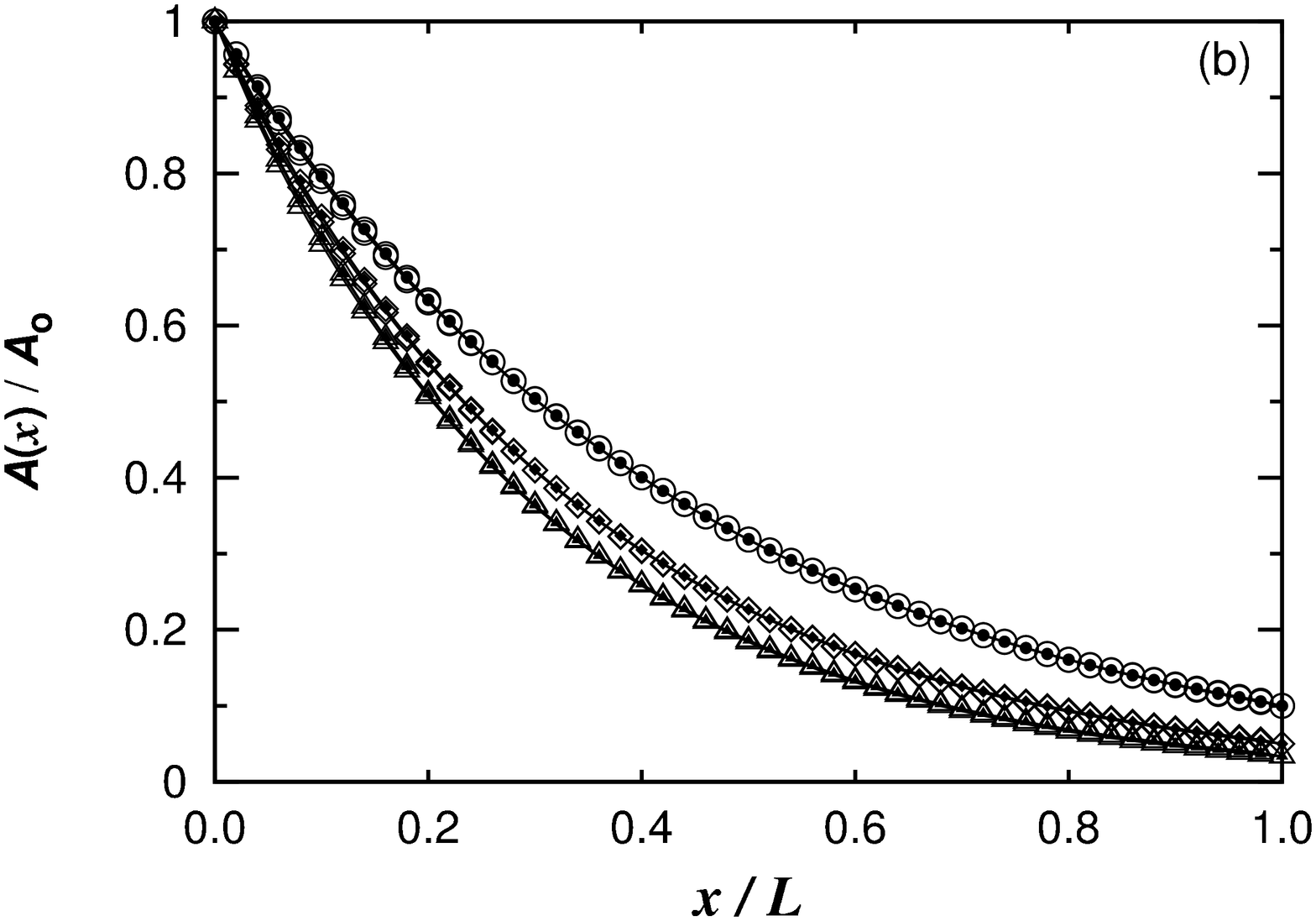}
\end{center}
\caption{
Cross section area as a function of position on 
the spinning line, at $t$=10,
for {\sf Dr}=10 (circle), 20 (diamond), 30 (triangle) and 
{\sf De}=10$^{-3}$ (open symbol), 10$^{-2}$ (filled symbol)
at (a) {\sf Re}=10$^{-2}$ and (b) {\sf Re}=0.1 and 1.
The solid lines (in (a) white and (b) black colors)
are the results obtained by using Maxwell constitutive equation 
with the parameters corresponding to the symbols.
}
\label{fig:Fig16}
\end{figure}
In Fig.\ref{fig:Fig16} we show 
the profiles of the cross section area 
at $t$=10 
for {\sf Dr}=10, 20, 30 and {\sf De}=10$^{-3}$, 10$^{-2}$
at (a) {\sf Re}=10$^{-2}$ and (b) {\sf Re}=10$^{-1}$ and 1.
All the profiles for {\sf Dr}=10 in (a) are on the same line,
{\it i.e., } they are independent of {\sf De}.
The profiles for {\sf Dr}=20 and 30, on the other hand, 
depend on {\sf De}, and the cross section areas for larger {\sf De} 
becomes thinner, particularly near the center region.
In (b), the area profiles depend only on the draw ratios, 
which is consistent with the results of the velocity profiles 
at {\sf Re}$\ge 10^{-1}$ shown in Fig.\ref{fig:Fig15}.
%
%
\begin{figure}[t]
\begin{center}
\includegraphics[scale=1,angle=-90,width=8.5truecm]{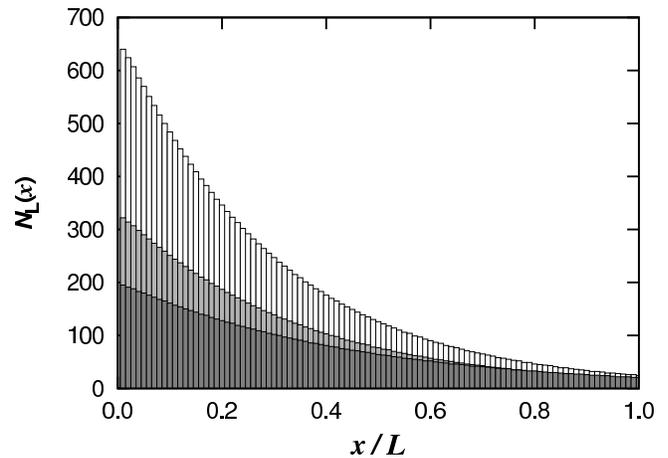}
\end{center}
\caption{
Typical histogram of the number of Lagrangian particles $N_\rL(x)$ 
existing in the spatial interval [$x,x+\D x$] along the spinning line 
at $t$=10, for {\sf Dr}=10 (dark grey), 20 (light grey), 30 (white).
}
\label{fig:Fig17}
\end{figure}
%
%
As with the low Reynolds number limiting case 
shown in Fig.\ref{fig:Fig06},
the distribution of Lagrangian particles on the spinning line
is a monotonically decreasing function of $x$.
This is because the density of particles 
is proportional to the inverse of the velocity. 
Since the stress tensor at a Eulerian grid point
is evaluated by averaging over the stress tensors 
of the Lagrangian particles located within a spatial interval of width
$\Delta x$, a larger number of particles will always 
increase the statistical accuracy and improve the numerical stability. 
In the present simulation, 
a new particle is inserted at 
$x$$\simeq$$0$, so that at least 
twenty Lagrangian particles always exist in the region closest to $x$=1,
as seen in Fig.\ref{fig:Fig17}.
The total numbers of Lagrangian particles in the steady state,
for {\sf Dr}=10, 20 and 30,
are $N_\rL^{\rm (Total)}\simeq$ 7850,  10600 and 19100, respectively.
%
%
\begin{figure}[t]
\begin{center}
\includegraphics[scale=1,width=3.2truecm,angle=-90]{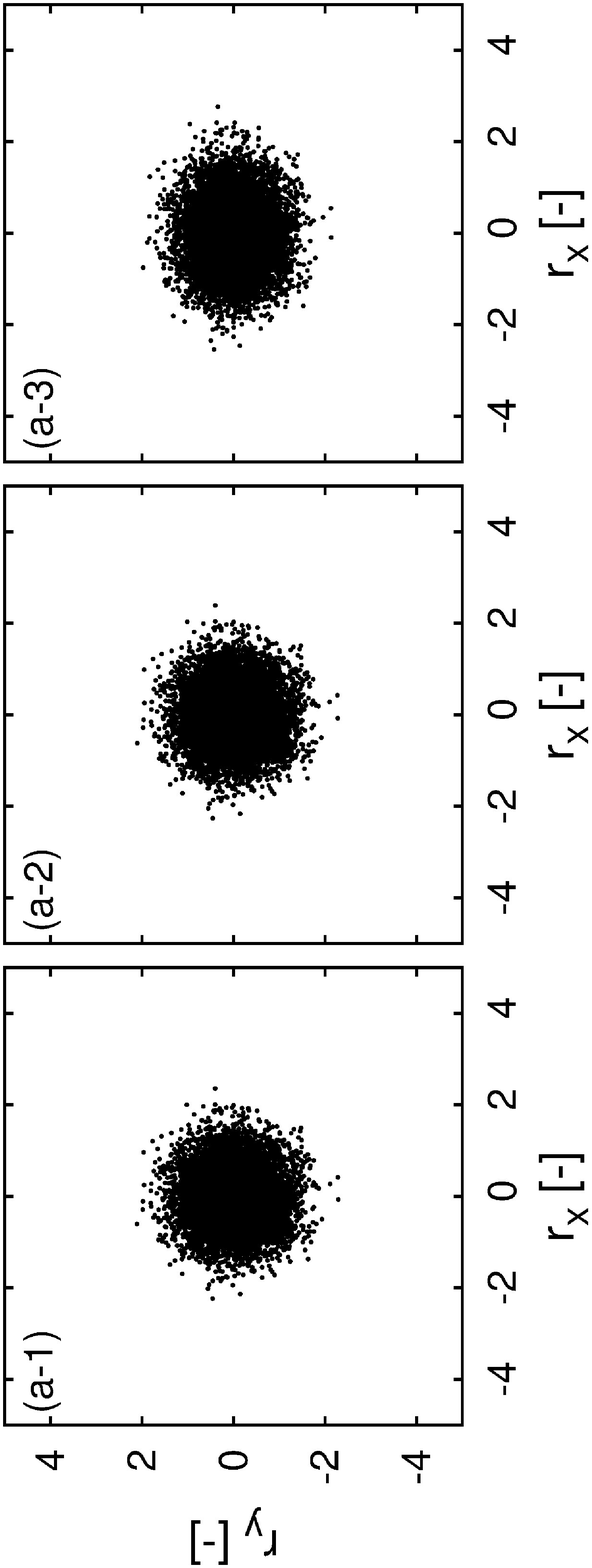}
\includegraphics[scale=1,width=3.2truecm,angle=-90]{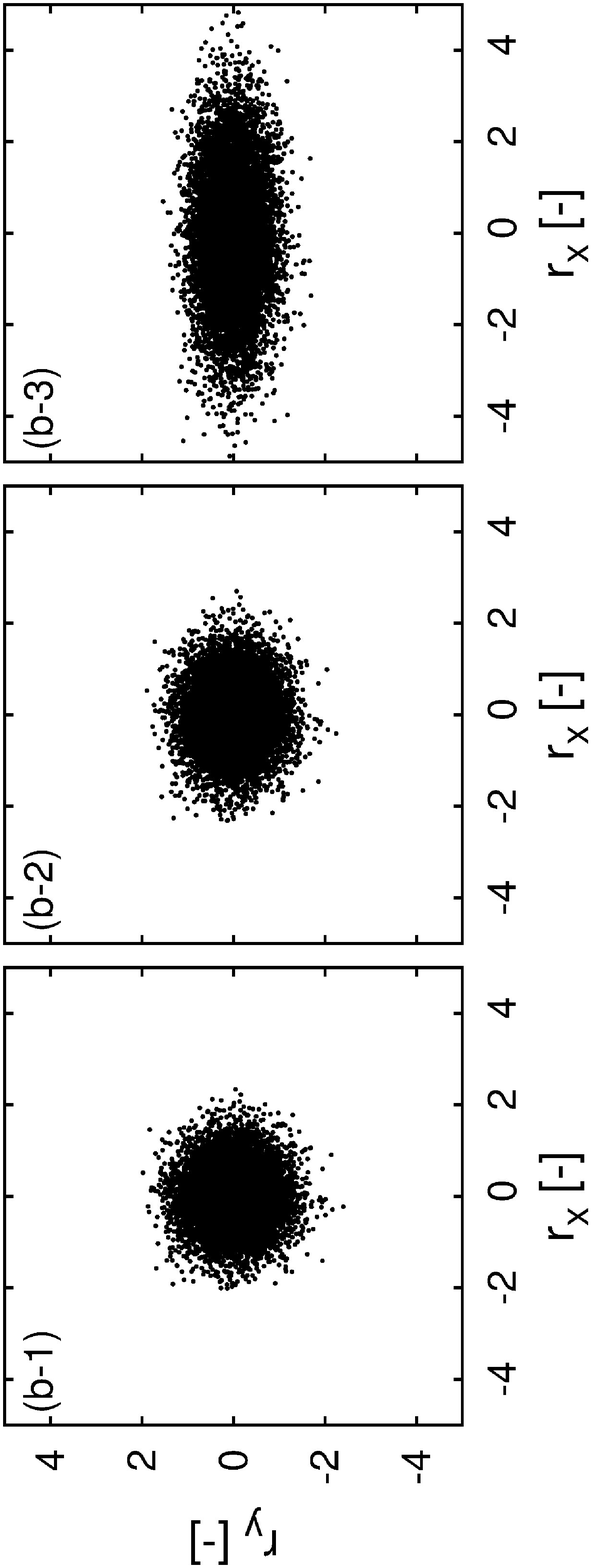}
\end{center}
\caption{
Distribution of connecting vectors 
in a Lagrangian particle located at (1) $x \simeq$ 0.1, (2) 0.5 and (3) 0.9
for 
(a) {\sf De}=10$^{-3}$ and 
(b) {\sf De}=10$^{-2}$ in ${\sf Dr}=30$ and {\sf Re}=1.
}
\label{fig:Fig18}
\end{figure}
%
%

As with Fig.\ref{fig:Fig10},
in Fig.\ref{fig:Fig18} we show the distribution 
of one of the two end points for dumbbells belonging
to  Lagrangian points located at 
(1) $x$$\simeq$0.1 
(2) $x$$\simeq$0.5
and 
(3) $x$$\simeq$0.9,
at $t$=10, for {\sf Re}=1 and {\sf Dr}=30
at (a) {\sf De}=10$^{-3}$ and (b) 10$^{-2}$.
In (a) {\sf De}=10$^{-3}$,  
the state of the dumbbells at $x$$\simeq$0.1 and 0.5 are almost isotropic,  
but the state at $x$$\simeq$0.9 is slightly stretched
along the flow direction.
In (b) {\sf De}=10$^{-2}$, on the other hand,
one can see that the state of the dumbbells is deformed from the isotropic state
even at $x$$\simeq$0.1, and at $x$$\simeq$0.9  
the dumbbells are highly stretched along $x$-direction and 
compressed in the $y$-direction, as seen in (b-3). 
%
%
%
\begin{figure}[t]
\begin{center}
\includegraphics[scale=1,width=3.2truecm,angle=-90]{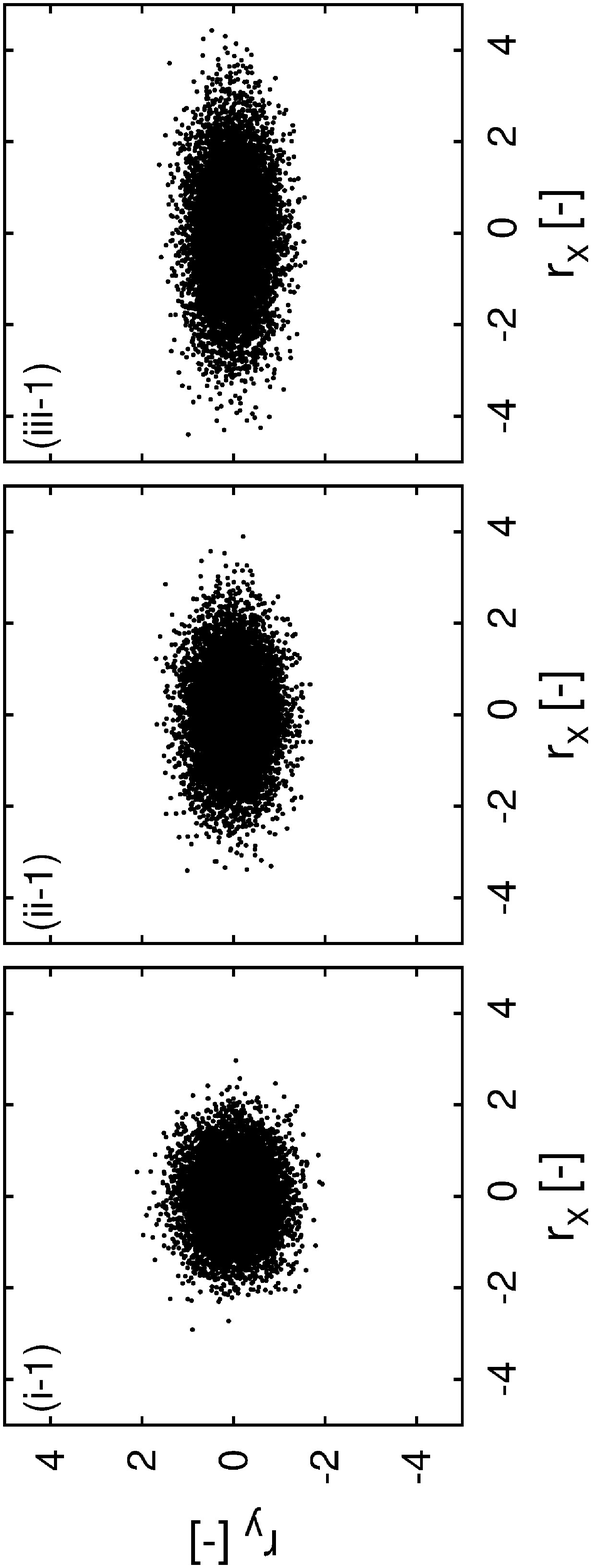}
\includegraphics[scale=1,width=3.2truecm,angle=-90]{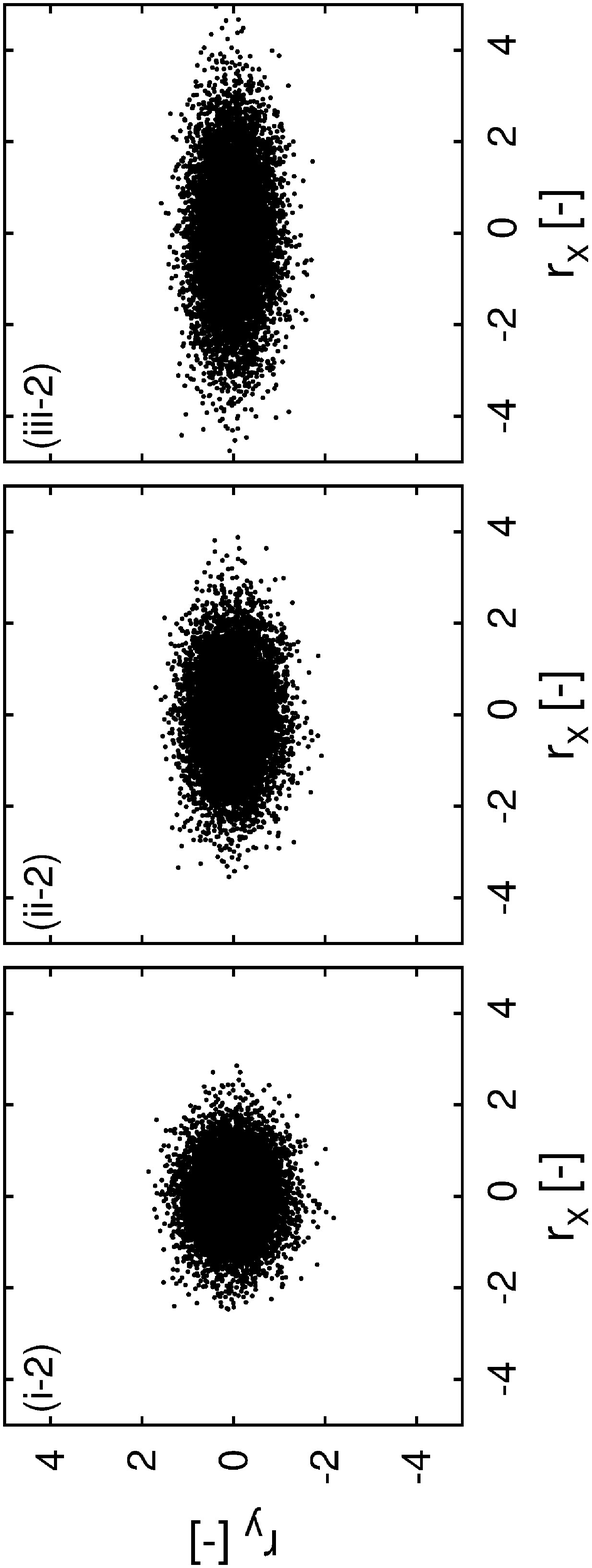}
\includegraphics[scale=1,width=3.2truecm,angle=-90]{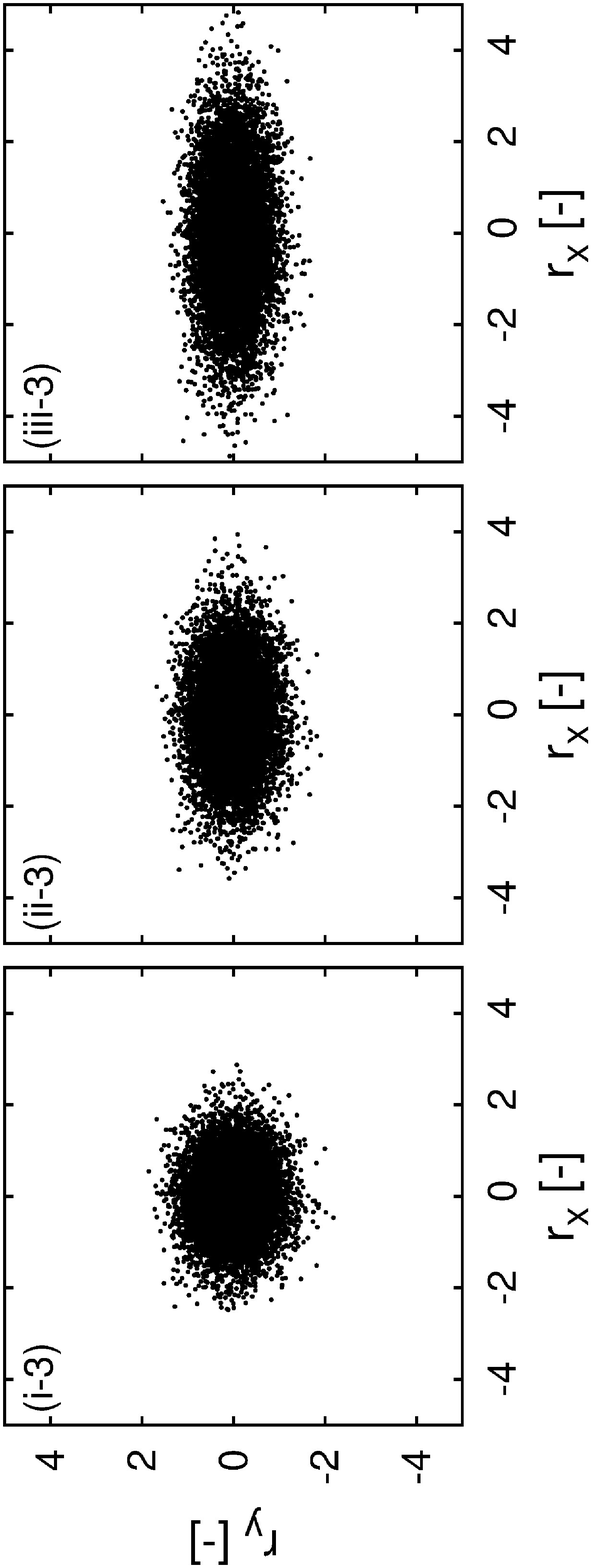}
\end{center}
\caption{
Distribution of connecting vectors
in a Lagrangian particle 
at $x$$\simeq$0.9 and at $t$=10
for 
(i)  {\sf Re}=10$^{-2}$, 
(ii) {\sf Re}=10$^{-1}$, 
and 
(iii) {\sf Re}=1, 
and 
(1) \Dr=10
(2) \Dr=20
and
(3) \Dr=30
in {\sf De}=10$^{-2}$.
}
\label{fig:Fig19}
\end{figure}
%
%
Fig.\ref{fig:Fig19} shows
the distribution of connecting vectors 
in a Lagrangian particle located at $x$$\simeq$0.9 
at $t$=10 
for the nine combinations of
{\sf Re}=10$^{-2}$, 10$^{-1}$, 1 and 
{\sf Dr}=10, 20, 30 in {\sf De}=10$^{-2}$.
The distributions are almost the same if the draw ratio {\sf Dr} is
the same, except for ({\sf i-1}) {\sf Re}=10$^{-2}$ and {\sf Dr}=10.
This result also means that 
the state of the dumbbells 
is determined only by {\sf Dr} if {\sf Re} is large enough, 
which is consistent with the results shown in 
Figs.\ref{fig:Fig15} and \ref{fig:Fig16},
where the velocity and cross section area 
converge towards profiles depending only on {\sf Dr}. 

\section{Summary}
  \label{sec:Summary}

   \medskip
   In the present study, 
we applied a multiscale simulation method 
to a melt spinning processes. 
In the simulations, we use a dumbbell model as the microscopic model
 for the polymer chains in order to validate our multiscale simulation method.
Based on the fact that 
the stress evaluated by the statistical average 
of a set of dumbbells is mathematically equivalent to the one calculated 
from the Maxwell constitutive equation,
we can directly assess the accuracy of our multiscale simulation method,
by comparing our results for the microscopic model,
with those obtained using the corresponding constitutive equation.
Actually, the validation of the multiscale simulation method 
is of particular importance, because the method 
itself is still under development.
The schemes used to solve the momentum equation
are chosen depending on the Reynolds number, {\it i.e.,}
(a) {\sf Re}$\rightarrow$0 and (b) finite {\sf Re}.
In the limit (a), the tension balance always holds
along the spinning line,
and the Broyden method is used to evaluate the velocity. 
We found that the elastic effect and a longer relaxation time
improve the stability of the spinning process, 
and that 
the draw resonance phenomenon that takes place 
above {\sf Dr}$\simeq$20.21
in the Newtonian fluid, is suppressed at {\sf De}=10$^{-2}$,
at least up to {\sf Dr}=30. 
In the limiting case in (b), on the other hand, 
the inertia effect starts to play an important role in determining 
the dynamical behavior of the spinning process
above \Re$\simeq$$10^{-1}$.
In this {\sf Re}-regime, 
the spinning process is quite stable, 
and the profiles for the cross section area and velocity 
along the spinning line are determined only by the draw ratio {\sf Dr}.
In both cases (a) and (b), 
we used $N_\rp$-dumbbells ($N_\rp$=10$^4$) in a Lagrangian particle. 
The use of a finite number of dumbbells in a Lagrangian particle
always brings with it a non-zero statistical error for the computed
stress, as compared to the one obtained by the Maxwell constitutive equation.
Even though the number of dumbbells in a Lagrangian particle is $N_\rp$=10$^4$
in this work, the present scheme gives good agreement with 
the results obtained by the Maxwell constitutive equation. 
In addition, we demonstrated that the MSS method 
is useful in bridging the macroscopic behavior and 
the simultaneous microscopic state,
by comparing the velocity and cross section area profiles with  
the end point distribution of the connecting vectors 
for the dumbbells in a Lagrangian particle at characteristic locations
along the spinning line.

We hope that the MSS method we propose brings us to a new stage in the
field of material science, by allowing us to develop novel approaches in
designing materials. In particular, we hope to increase the current
understanding of the flow behaviors of polymeric materials, 
not only from a macroscopic point of view, but also from a detailed microscopic level.

In the near future, 
by employing a more realistic microscopic model for the polymer chains, 
such as the
primitive chain network models 
({PASTA\cite{PASTA_Doi_Takimoto2001,Doi2003}},
{NAPLES\cite{NAPLES_Masubuchi2001}}),
an improved description of the entanglements of the polymer chains and
their orientations will be incorporated into our MSS method.
In the present work, we have simply used an isothermal condition. 
However, in many industrial problems,
the temperature is not homogeneous and 
a solidification and/or a crystallization of the polymeric material 
may occur.
To deal with such phenomena, we need to extend the microscopic model
or even develop a new one.
In addition, to overcome the problems due to the high computational cost,
we will have to improve the numerical schemes,
including the parallelization of the numerical codes.
Thereby, the MSS method will be more effective
in solving problems with industrial applications.

The authors would like to express our sincere gratitude
to J. Molina for giving valuable comments 
and for kindly proof-reading the present paper.
This work was supported partially by JSPS KAKENHI Grant Number 15K13549 and 
by the FLAGSHIP2020, MEXT within the challenging study, sub-theme B.



\appendix

\section{Analytic Solutions of steady state at limits}


We need to rely on a numerical method 
to obtain a steady state solution of
eqs.\eqref{eqn:rescaled_eq_of_continuity}-%
\eqref{eqn:rescaled_BoundaryCondition_to_V_at_x=1}
and \eqref{eqn:dimensionless_constitutive_eq_for_dumbbell_model}
at a finite Deborah number,
but in the two limits: 
(I) \De$\rightarrow$$0$ and (II) \De$\rightarrow$$\infty$,
one can obtain analytic solutions for the steady states.

\medskip
\noindent
{(I) {\sf De}$\rightarrow 0$ {\rm (Newtonian limit)}}

\noindent
The analytic solution of the steady state 
for the spinning process of a Newtonian fluid at \Re$\ne$$0$ 
can be obtained as
\begin{equation}
V(x)=\Dr { \Re - (\Re - C )e^{C/3} 
          \over 
           \Re - (\Re - C )e^{Cx/3}
          }, 
\quad
A(x)=[V(x)]^{-1}
\end{equation}
with the constant $C$ determined by the following equation:
\begin{equation}
C + \Dr (\Re - C )e^{C/3} = \Dr \Re .
\end{equation}
The tension $F$ is given by 
\begin{equation}
F(x) = {C  (\Re - C )e^{Cx/3} \over \Re - (\Re - C )e^{Cx/3} }.
\end{equation}
If one takes the limit \Re$\rightarrow$$0$,
the constant $C$ is found to be $-3\ln(\Dr)$, 
and the above analytic solution turns out be 
the well-known analytic solution 
at the steady state ({\sf Dr}$<$20.21)
\cite{Gelder1971,Fisher1975}
of the spinning process for a Newtonian fluid with \Re=0 as
\begin{equation}
V(x)=({\sf Dr})^{x}, 
\quad 
A(x)=({\sf Dr})^{-x}, 
\quad
F   = 3 \ln({\sf Dr}).
\label{eqn:AnalyticSolution_of_Newtonian_fluid}
\end{equation}

\medskip
\noindent
{(II) {\sf De}$\rightarrow \infty$ {\rm (Elastic limit)}}

\noindent
In the elastic limit $(${\sf De}$\rightarrow\infty)$,
one can also obtain the analytic solution as
\begin{eqnarray}
&& \hspace{-13mm} 
   V(x)= 1 + ( {\sf Dr} - 1 )x, \quad 
   A(x)= \Bigr [1 + ( {\sf Dr} - 1 )x \Bigr ]^{-1}.
\label{eqn:analytic_solution_in_elastic_limit}   
\end{eqnarray}
The elastic stresses for $xx$-, $yy$- and $zz$-components are 
\begin{eqnarray}
&& \hspace{-13mm} 
   \sigma_{xx} = \Re \Dr^2 {\Dr-1 \over \Dr^3-1} [V(x)]^2   \\
&& \hspace{-13mm} 
   \sigma_{yy} = \sigma_{yy} = {1\over 2}
                 \Re \Dr^2 {\Dr-1 \over \Dr^3-1} [V(x)]^{-1}.
\end{eqnarray}
Then the tension is given as
\begin{eqnarray}
F = \Re \Dr^2 {\Dr-1 \over \Dr^3-1} \Bigr [ V(x) - V(x)^{-2} \Bigr ]. 
\end{eqnarray}



%

%


\begin{thebibliography}{100}
\bibitem{Ishihara_2011_1} 
H. Ishihara, {\it Seikei-Kakou}, {\bf 23}, 276-285 (2011)
{\it in Japanese}. 
\bibitem{Ishihara_2011_2} 
H. Ishihara, {\it Seikei-Kakou}, {\bf 23}, 336-346 (2011)
{\it in Japanese}. 
\bibitem{Ishihara_2011_3} 
H. Ishihara, {\it Seikei-Kakou}, {\bf 23}, 430-440 (2011)
{\it in Japanese}. 
\bibitem{KaseMatsuo1965}
S. Kase and T. Matsuo, {\it J. Polym. Sci. Part A}, {\bf 3}, 2541 (1965).
\bibitem{KaseMatsuo1967}
S. Kase and T. Matsuo, {\it J. Polym. Sci. Part A}, {\bf 11}, 251 (1967).
\bibitem{MatovichPearson1969}
M. A. Matovich and J. R. A Pearson,
{\it Ind. Eng. Chem. Fund.}, {\bf 8}, 512-520 (1969). 
\bibitem{PearsonMatovich1969}
J. R. A. Pearson and M. A. Matovich, 
{\it Ind. Eng. Chem. Fund.}, {\bf 8}, 606 (1969).
\bibitem{Gelder1971}
D. Gelder,
{\it Ind. Eng. Chem. Fund.}, {\bf 10}, 534 (1971).
\bibitem{Shah1972}
%
Y. T. Shah and J.R.A.Pearson, {\it Ind.Eng. Chem. Fund.} {\bf 11} 150, (1972).
\bibitem{Ishihara1973}
H. Ishihara and S. Hayashi, 
{\it Nihon Reoroji Gakkaishi}, {\bf 17}, 19 (1973)
{\it in Japanese}. 
\bibitem{Denn1975}
M. M. Denn and C. J. S. Petrie, 
{\it AIChE J.}, {\bf 21}, 791 (1975).
\bibitem{Fisher1975}
R. J. Fisher and M. M. Denn,
{\it Chem. Eng. Sci.}, {\bf 30}, 1129  (1975)
\bibitem{Ishihara1976}
H. Ishihara, {\it J. Appl. Polym. Sci.}, {\bf 20}, 169 (1976).
\bibitem{Hyun1978b}
J. C. Hyun, {\it AIChE J.} {\bf 24}, 423 (1978).
\bibitem{Ishihara_1989}  
H. Ishihara and S. Hayashi, 
{\it Nihon Reoroji Gakkaishi}, {\bf 17}, 19 (1989)
\bibitem{Ishihara_1992a} 
H. Ishihara, S. Hayashi and H. Yasuda, 
{\it Nihon Reoroji Gakkaishi}, {\bf 20}, 109 (1992), {\it in Japanese}.
\bibitem{Ishihara2006}
H. Ishihara, M. Shibata and K. Ikeda, 
{\it Nihon Reoroji Gakkaishi}, {\bf 34}, 3 (2006), {\it in Japanese}.
\bibitem{Takarada2001}
W. Takarada {\it et al}, 
{\it Journal of Applied Polymer Science}, {\bf 80}, 1589 (2001).
\bibitem{Yun2008}
J. H. Yun, {\it Nihon Reoroji Gakkaishi}, {\bf 36} 133 (2008).
\bibitem{Perera2008}
S. S. N. Perera, {\it Nihon Reoroji Gakkaishi}, {\bf 36}, 161 (2008).
\bibitem{Dhadwal2011}
R. Dhadwal, {\it Applied Mathematical Modelling}, {\bf 35}, 2959 (2011).
\bibitem{Rouse1953}
P. E. Rouse, {\it J. Chem. Phys.}, {\bf 21}, 1272 (1953).
\bibitem{book:Bird1987}
R. B. Bird, R. C. Armstrong, and O. Hassager,
{\it Dynamics of polymeric liquids}
Vol. 1 and 2 (John Wiley \& Sons, New York, 1987).
\bibitem{KremerGrest1990}
K. Kremer and G. S. Grest,
{\it J. Chem. Phys.}, {\bf 92}, 5057 (1990).
\bibitem{DE1986}
M. Doi and S. F. Edwards,
{\it The theory of polymer dynamics.}, 
(Oxford University Press, New York, 1986).
\bibitem{NAPLES_Masubuchi2001}
Y. Masubuchi {\it et al},
{\it J. Chem. Phys.}, {\bf 115}, 4387 (2001).
\bibitem{ShanbhagLarson2001}
S. Shanbhag {\it et al},
{\it Phys. Rev. Lett.}, {\bf 87}, 195502 (2001).
\bibitem{PASTA_Doi_Takimoto2001}
H. Tasaki, J. Takimoto and M. Doi,
{\it Comput. Phys. Commun.}, {\bf 142}, 136, (2001).
\bibitem{Doi2003} 
M. Doi and J. Takimoto, 
{\it Phil. Trans. R. Soc. Lond.}, {\bf A 361}, 641 (2003).
\bibitem{Schieber2003}
J. D. Schieber, J. Neergaard, and S. Gupta,
{\it J. Rheol.}, {\bf 47}, 213 (2003).
\bibitem{Likhtman2005}
A. E. Likhtman,
{\it Macromolecules}, {\bf 38}, 6128 (2005).
%
\bibitem{Ottinger2005}
H. C. {\"O}ttinger,
{\it Beyond Equilibrium Thermodynamics}  
(Wiley-Interscience, New Jersey, 2005).
\bibitem{LasoOttinger1993}
M. Laso and H. C. \"Ottinger,
{\it J. Non-Newtonian Fluid Mech.}, {\bf 47}, 1 (1993).
\bibitem{Feigl1995}
K. Feigl, M. Laso, and H. C. \"Ottinger,
{\it Macromolecules}, {\bf 28}, 3261 (1995).
\bibitem{WedgewoodGeurts1995}
L. E. Wedgewood and K. R. Geurts, 
{\it Ind. Eng. Chem. Res.}, {\bf 34}, 3437 (1995).
\bibitem{HuaSchieber1996}
C. C. Hua and J. D. Schieber,
{\it Chem. Eng. Sci.}, {\bf 51}, 1473 (1996).
\bibitem{vanHeelHulsenBrule1998}
A.P.G. van Heel {\it et al}, 
{\it J. Non-Newtonian Fluid Mech.}, {\bf 75}, 253 (1998).
\bibitem{LasoPicassoOttinger1997}
M. Laso, M. Picasso and H. C. \"Ottinger,
{\it AIChE J.}, {\bf 43}, 877 (1997).
\bibitem{HalinLielensKeuningsLegat1998}
P. Halin {\it et al}, 
{\it J. Non-Newtonian Fluid Mech.}, {\bf 79}, 387 (1998) 
\bibitem{BonvinPicasso1999}
J. Bonvin and M. Picasso,
{\it J. Non-Newtonian Fluid Mech.}, {\bf 84}, 191 (1999) 
\bibitem{RenE2005}
W. Ren and W. E,
{\it J. Compt. Phys.}, {\bf 204}, 1 (2005).
\bibitem{WeinanE2007}
W. E, {\it et al}, 
{\it Commun. Comput. Phys.}, {\bf 2}, 367 (2007).
\bibitem{YasudaYamamoto2008}
S. Yasuda and R. Yamamoto,
{\it Phys. Fluids}, {\bf 20}, 113101 (2008).
\bibitem{YasudaYamamoto2009}
S. Yasuda and R. Yamamoto,
{\it Europhys. Lett.}, {\bf 86}, 18002 (2009).
\bibitem{YasudaYamamoto2010}
S. Yasuda and R. Yamamoto,
{\it Phys. Rev. E}, {\bf 81}, 036308 (2010).
\bibitem{YasudaYamamoto2011}
S. Yasuda and R. Yamamoto,
{\it Phys. Rev. E}, {\bf 84}, 031501 (2011).
\bibitem{MurashimaTaniguchi2010}
T. Murashima and T. Taniguchi,
{\it J. Polym. Sci. B}, {\bf 48}, 886 (2010).
\bibitem{MurashimaTaniguchi2011}
T. Murashima and T. Taniguchi,
{\it Europhys. Lett.}, {\bf 96}, 18002 (2011).
\bibitem{MurashimaTaniguchi2011b}
T. Murashima and T. Taniguchi,
{\it J. Phys. Soc. Jpn.}, SA013 (2011).
\bibitem{Murashima2013}
T. Murashima {\it et al}, 
{\it J. Phys. Soc. Jpn.}, {\bf 82}, 012001 (2013).
\bibitem{Isothermal_Assumption}
If one wants to simulate industrial melt spinning processes,
the isothermal assumption should be replaced by a non-isothermal one,
but we are considering that the ideal isothermal condition is suitable
to make an assessment of the present multiscale simulation.
\bibitem{t_ast}
The time period $t^\ast$ is estimated by 
$t^\ast$=$\int_0^1 dx/V(x)$ and $V(x)$
in \eqref{eqn:AnalyticSolution_of_Newtonian_fluid}.
\bibitem{CPU}
In the present works, 
(i)  24Cores: Xeon E5-1680v3 (3.2GHz) and
(ii) 16Cores: Xeon E5-2680 (2.7GHz) are used in (a),
and the CPU (i) is used in (b).
\end{thebibliography}
                            \end{document}